\documentclass[10pt,journal,compsoc]{IEEEtran}

\usepackage{amsmath,amsfonts}
\usepackage{algorithmic}
\usepackage{algorithm}
\usepackage{array}
\usepackage[caption=false,font=normalsize,labelfont=sf,textfont=sf]{subfig}
\usepackage{textcomp}
\usepackage{stfloats}
\usepackage{url}
\usepackage{verbatim}
\usepackage{graphicx}
\usepackage{orcidlink}
\usepackage{cite}
\usepackage{utils}

\begin{document}

\title{\newrevised{RECOVER: Toward Requirements
Generation from Stakeholders' Conversations}}

\author{Gianmario Voria\orcidlink{0009-0002-5394-8148}, Francesco Casillo\orcidlink{0000-0003-4869-8068}, Carmine Gravino\orcidlink{0000-0002-4394-9035}, Gemma Catolino\orcidlink{0000-0002-4689-3401}, Fabio Palomba\orcidlink{0000-0001-9337-5116}
\IEEEcompsocitemizethanks{\IEEEcompsocthanksitem Gianmario, Francesco, Carmine, Gemma, and Fabio are with the Software Engineering (SeSa) Lab of the University of Salerno, Italy.
\protect\\
E-mails: \{gvoria, fcasillo, gravino, gcatolino, fpalomba\}@unisa.it
}
}

\markboth{IEEE Transactions on Software Engineering}%
{Voria \MakeLowercase{et al.}: Requirements Generation from Stakeholders' Conversations}

\IEEEtitleabstractindextext{%
\begin{abstract}
Stakeholders' conversations in requirements elicitation meetings hold valuable insights into system and client needs. However, manually extracting requirements is time-consuming, labor-intensive, and prone to errors and biases. While current state-of-the-art methods assist in summarizing stakeholder conversations and classifying requirements based on their nature, there is a noticeable lack of approaches capable of both identifying requirements within these conversations and generating corresponding system requirements. These approaches would assist requirement identification, reducing engineers' workload, time, and effort. They would also enhance accuracy and consistency in documentation, providing a reliable foundation for further analysis. To address this gap, this paper introduces \textsc{RECOVER} (Requirements EliCitation frOm conVERsations), a novel conversational requirements engineering approach that leverages natural language processing and large language models (LLMs) \revised{to support practitioners in automatically extracting} system requirements from stakeholder interactions by analyzing individual conversation turns. The approach is evaluated using a mixed-method research design that combines statistical performance analysis with a user study involving requirements engineers, targeting two levels of granularity. First, at the conversation turn level, the evaluation measures \textsc{RECOVER}'s accuracy in identifying requirements-relevant dialogue and the quality of generated requirements in terms of correctness, completeness, and actionability. Second, at the entire conversation level, the evaluation assesses the overall usefulness and effectiveness of \textsc{RECOVER} in synthesizing comprehensive system requirements from full stakeholder discussions. Empirical evaluation of \textsc{RECOVER} shows promising performance, with generated requirements demonstrating satisfactory correctness, completeness, and actionability. The results also highlight the potential of automating requirements elicitation from conversations \revised{as an aid that enhances efficiency while maintaining human oversight}.
\end{abstract}

}

\maketitle
\IEEEpeerreviewmaketitle

\section{Introduction}
Requirements Engineering revolves around the elicitation, analysis, and specification of functional and non-functional requirements that a software system must guarantee to its users \cite{bruegge1999object,zowghi2005requirements}. Research has shown that key information for eliciting requirements is often gathered through conversations between requirements engineers and system stakeholders \cite{davis2006effectiveness}. These interactions typically occur during interviews or moderated workshops, which are increasingly employed to gain a clearer understanding of the features a software system should incorporate \cite{wagner2019status}.

Recent studies highlight that stakeholder conversations aids in early detection of ambiguity and misconceptions in requirements engineering \cite{ferrari2016ambiguity, stolcke2000dialogue}. However, this process demands significant cognitive effort and time \cite{reconsum_spijkman_2023}, with agreement taking from hours to weeks \cite{alvarez2002tell, spijkman2021requirements}. Note-taking may work for short exchanges, but longer discussions risk missing key details, reducing elicitation effectiveness \cite{spoletini2018interview}. Additionally, reviewing recordings is tedious and error-prone \cite{reconsum_spijkman_2023}, particularly in critical systems like ML-enabled projects and system-of-systems, where implicit requirements (e.g., fairness, privacy) may be overlooked \cite{katina2014system, nahar2022collaboration}.


To address these challenges, Spijkman et al. \cite{reconsum_spijkman_2023} recently proposed \textsc{REConSum}, an automated approach that leverages natural language processing to (1) summarize conversations and (2) detect requirements-relevant questions and answers. By ``requirements-relevant'', the authors referred to exchanges that contain information potentially useful for identifying system requirements, e.g., discussions about user needs, specific system functionalities, constraints, or technical specifications. According to the insights provided by Spijkman et al. \cite{reconsum_spijkman_2023}, \textsc{REConSum} reduces the effort required by requirements engineers in eliciting requirements from conversations, streamlining the process by focusing on the most pertinent information.

Recognizing the seminal advances made by \textsc{REConSum}, this paper performs a further step ahead by proposing a three-step approach---coined \textsc{RECOVER} (\textbf{R}equirements \textbf{E}li\textbf{C}itation fr\textbf{O}m con\textbf{VER}sations)---that can \revised{assist requirements engineers in systematically identifying and structuring system requirements from conversations}. Rather tthan summarizing the input conversations, \textsc{RECOVER} considers the entire corpus of the conversation to individually classify requirements-relevant conversation turns. It then processes these turns to filter out noise and enhance them with contextual information. Next, \textsc{RECOVER} exploits the capabilities of large language models (LLMs) to generate system requirements corresponding to the identified requirements-relevant conversation turns. Finally, the approach aggregates the requirements generated for each relevant turn, providing requirements engineers with the full set of system requirements generated. \revised{While \textsc{RECOVER} leverages automation to support the elicitation process, it is not intended as a fully autonomous system but rather as an aid that enhances efficiency while maintaining human oversight.}

We evaluate \textsc{RECOVER} through an empirical study using the conversation dataset provided by Spijkman et al. \cite{reconsum_spijkman_2023}. This study includes a statistical performance analysis and a user study with 23 experienced requirements engineers, examining two granularity levels. At the \textbf{individual conversation turn} level, we assess \textsc{RECOVER}'s accuracy in classifying requirements-relevant turns, its ability to generate system requirements matching those manually produced by engineers, and the correctness, completeness, and actionability of the generated requirements. At the \textbf{entire conversation} level, we assess \textsc{RECOVER}'s overall usefulness by examining whether focusing on individual turns omits relevant requirements. This evaluation compares \textsc{RECOVER}'s output against two baselines: one produced by requirements engineers and another generated by ChatGPT, a large language model that does not preprocess requirements-relevant conversation turns. The inclusion of an LLM baseline aims at demonstrating the added value of \textsc{RECOVER}'s targeted preprocessing capabilities with respect to a more general approach that lacks the same level of contextual understanding and specificity.

At the \textbf{individual conversation turn granularity}, \textsc{RECOVER} may classify requirements with a recall of 76\%. \revised{In addition, the requirements generated by \textsc{RECOVER} have lower BLEU scores (Mean 5.39\%) when considering precise wording. At the same time, they can properly capture content and semantics from conversations, as outlined by METEOR scores (Mean 41.87\%) and ROUGE (Mean 38.53\%)}. Finally, requirements engineers found the approach valuable, yet they still raised the need for post-process validations conducted by experts to remove erroneous requirements produced by hallucinations of LLMs.

At the \textbf{entire conversation granularity}, the high BLEU (78.82\%) and METEOR (39.09\%) scores achieved by \textsc{RECOVER} when compared against the manual baseline indicate its strong ability to generate requirements that are both accurate and comprehensive, making it a reliable tool for producing outputs closely aligned with expert standards. In contrast, while ChatGPT's higher ROUGE score (43.96\%) suggests it captures a broad range of content similar to the manual baseline, its lower BLEU score highlights potential shortcomings in precision and comprehensiveness. This comparison underscores that utilizing the full \textsc{RECOVER} approach, rather than relying solely on a language model prompt, yields better results.

To sum up, our paper provides three major contributions to the field of requirements engineering:

\begin{enumerate}
    \item A novel, automated approach to recover requirements from stakeholders' conversations, which \revised{is a valuable tool for supporting practitioners during the} requirements elicitation processes;

    \smallskip
    \item A mixed-method empirical assessment of the approach, which measured the performance and overall quality of \textsc{RECOVER} at two different levels of granularity;

    \smallskip
    \item \newrevised{A preliminary in-vivo evaluation of RECOVER on industrial elicitation transcripts to assess its performance in realistic, noisy settings and its applicability across different conversational contexts;}

    \smallskip
    \item A publicly available replication packages \cite{appendix}, that researchers may use to either replicate/reproduce our results or build on top of them. 
    
\end{enumerate}

\section{Related Work and Contribution}
\label{sec:related}

Our work intersects multiple areas within Requirements Engineering, focusing on using conversational artifacts for requirement generation. Our approach, \textsc{RECOVER}, distinguishes itself by employing machine learning techniques to identify potential requirements within conversation excerpts and leveraging large language models (LLMs) to generate detailed requirements from requirements-relevant conversation turns. Jones's early work \cite{jones2004software} highlighted the challenges in managing software projects, emphasizing the importance of clear and comprehensive requirements elicitation, which aligns with our goal to enhance and streamline the requirements generation process.

Recent studies have explicitly focused on requirements conversations, analyzing their characteristics. Ferrari et al. \cite{ferrari2016ambiguity, ferrari2021using} examined ambiguity in requirements interviews and incorporated voice and biofeedback into conversational artifacts. Our approach differs by applying automated solutions to analyze these conversations, offering a more nuanced understanding and extraction of requirements. Alvarez and Urza \cite{alvarez2002tell} explored stakeholder roles through interview transcripts, providing insights into the complexities of requirements discussions. Our research builds on their work by automating the identification of pertinent information in similar transcripts, thereby enhancing the efficiency and accuracy of extracting requirements.

Pre-requirements specification traceability, initially highlighted by Gotel and Finkelstein \cite{Gotel1994AnAO} to differentiate pre- and post-requirements, was further explored by Krause et al. \cite{krause2020code} through a qualitative analysis of 67 papers, identifying its significance and challenges. Our methodology aligns with this foundation, aiming to enhance requirements management using systematic analysis and machine learning to trace requirements from early dialogues. \revised{However, while \textsc{RECOVER} inherently maintains an implicit connection between elicited requirements and their source dialogues by analyzing conversation turns, it does not explicitly support traceability. Instead, our approach conceptually aligns with pre-requirements specification traceability by bridging early-stage stakeholder discussions with structured requirements.} Similarly, van der Aa et al. \cite{vanderaa2019extracting} developed an automated method for deriving declarative process models from natural language, addressing challenges like synonyms and phrase discrepancies. A recent review \cite{hou2024largelanguagemodelssoftware} highlights LLMs' potential in software engineering, particularly in enhancing requirement clarity and reducing interpretational uncertainties. These insights are key to \textsc{RECOVER}, as its LLM component addresses linguistic variations.

Studies by Kurtanović and Maalej \cite{kurtanovic2017automatically} have provided automated methods for classifying requirements in specifications. \revised{While their work focuses on distinguishing functional from non-functional requirements in structured documents, our approach addresses a different but related challenge—identifying whether a conversation turn contains a requirement. Both tasks involve applying automated classification techniques to textual requirements artifacts, highlighting the broader and established role of machine learning in requirements classification against newer technologies, such as LLMs.} The proposed approach advances these methodologies by employing sophisticated AI-driven analysis to enhance the requirements generation process.



Finally, the work of Spijkman et al. \cite{spijkman2022back, spijkman2019specification, spijkman2021alignment, spijkman2021requirements} offered insights into requirements elicitation and documentation. Through multiple studies, the authors examined the field of conversational requirements engineering; they identified the challenges associated with manually extracting valuable information from conversations, citing the process as time-consuming and inefficient. Recognizing the need for automation, they developed \textsc{REConSum} \cite{reconsum_spijkman_2023}, an automated approach able to summarize requirement conversations and locate speaker turns---particularly questions---that may eventually contain requirements-relevant information. These studies align closely with the goals of \textsc{RECOVER}, which also seeks to improve the efficiency and accuracy of requirements elicitation and documentation by leveraging advanced machine learning and LLM technologies. \revised{However, while \textsc{RECOVER} builds upon the core intuitions of \textsc{REConSum}, it introduces multiple modifications. First, whereas \textsc{REConSum} only identifies speaker turns where requirements might be discussed, \textsc{RECOVER} extends this process to fully elicit system requirements using an LLM after the first two steps. Second, while \textsc{REConSum} assumes that requirement-relevant information is best extracted from Q\&A patterns, we found this approach limiting, as it may overlook cases where stakeholders directly assert requirements without a preceding question. To mitigate this potential cause of inaccuracy, we inverted the original steps: instead of first identifying Q\&A patterns and then filtering for requirement-relevant turns, \textsc{RECOVER} first classifies conversation turns to determine which contain requirement-related information. Only then does it apply the Q\&A-based structuring to enhance context and facilitate requirement generation. Additionally, unlike \textsc{REConSum}, which relies on a predefined set of domain-specific terms to determine requirement relevance, \textsc{RECOVER} employs a machine learning-based classifier. This shift was required due to the step inversion---since classification occurs before Q\&A tagging, relying on static term-based filtering would have been insufficient to capture relevant turns.}

\steSummaryBox{\faList \hspace{0.05cm} Summary and Contribution.}{\textsc{RECOVER} advances RE research by leveraging ML and LLMs to automate requirements extraction from conversations, enhancing efficiency and accuracy in elicitation. Our design integrates concepts from \textsc{REConSum} \cite{reconsum_spijkman_2023}, positioning \textsc{RECOVER} as a complementary work. While \textsc{REConSum} summarizes conversations to extract requirements-relevant information primarily from Q\&A structures, \textsc{RECOVER} aims to fully automate requirements elicitation by considering each conversation turn as potentially containing valuable information, \revised{supporting practitioners in the actual elicitation of system requirements rather than only suggesting potentially-relevant turns}.}

\begin{figure*}
    \centering
    \includegraphics[width=0.8\textwidth]{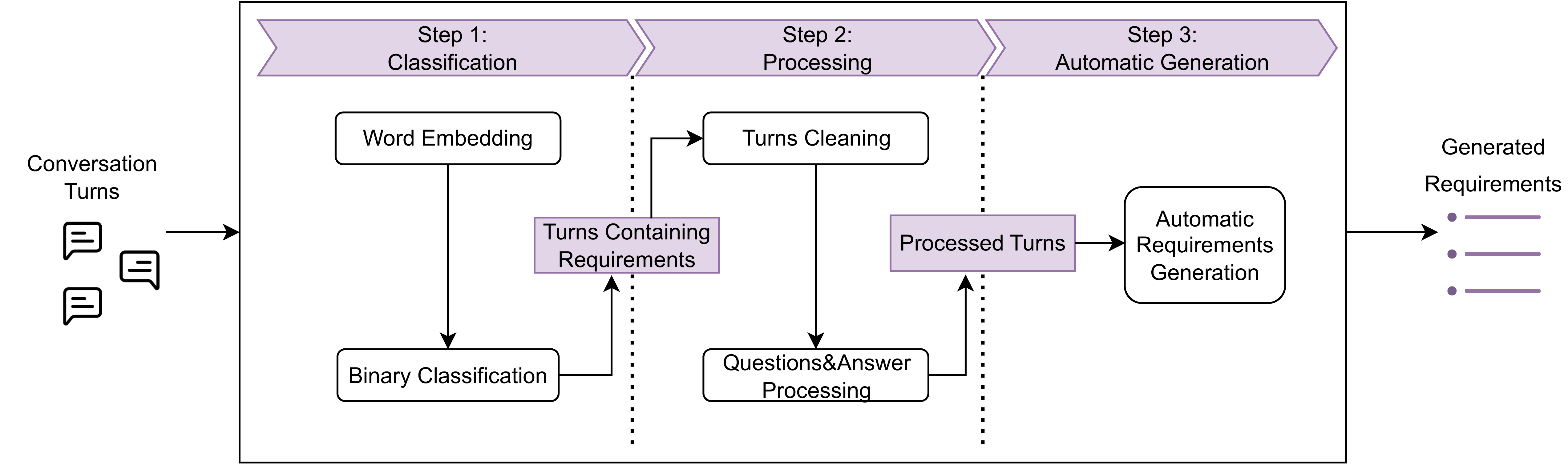}
    \caption{An overview of the main steps performed by \textsc{RECOVER}.}
    \label{figure:approach}
\end{figure*}

\section{RECOVER: A Novel Conversational Requirements Engineering Approach}
\label{sec:approach}
In this section, we introduce \textsc{RECOVER}, a novel approach to conversational requirements engineering designed to extract system requirements from unstructured stakeholder conversations. As illustrated in Figure \ref{figure:approach}, \textsc{RECOVER} operates through three main steps: (1) \textsl{`Classification'} of individual conversation turns to identify requirements-relevant content \revised{using shallow machine learning}, (2) \textsl{`Processing'} of the selected turns to identify and filter Question-Answer pairs \cite{reconsum_spijkman_2023}, discarding invalid ones, and (3) \textsl{`Automatic generation'} of system requirements from the processed conversation turns using a Large Language Model. \revised{\textsc{RECOVER} is designed to process stakeholder conversations in a generalizable manner, provided they are structured as a list of textual conversation turns. Unlike approaches that rely on predefined speaker roles, \textsc{RECOVER} is setting-agnostic, processing each conversation turn in isolation without explicitly considering the number or roles of the speakers.} \revised{In the context of our work, a \emph{conversation turn} refers to a set of contiguous utterances made by the same speaker before another participant takes over \cite{hepburn2012conversation}.} 

\revised{While many elicitation conversations follow a structured question-answer pattern, we found that relying solely on such structures could be limiting for requirement extraction. Therefore, \textsc{RECOVER} identifies requirements from both individual conversation turns and structured Q\&A sequences. To ensure completeness, Step \#2 of \textsc{RECOVER} detects turns with a question-like structure and pairs them with subsequent turns, assuming they contain corresponding answers. This approach minimizes the risk of missing relevant requirements, even at the cost of some redundancy}.

\revised{Our approach only leverages an LLM in the third step: while these models are promising, we decided against using them for the early steps of \textsc{RECOVER} due to concerns over consistency, accuracy, and study feasibility.
First, LLMs are inherently non-deterministic, meaning their outputs can vary unpredictably, even for the same input. \textsc{RECOVER} was designed as a structured, multi-step pipeline that builds upon prior work \cite{reconsum_spijkman_2023} while addressing its limitations. The potential lack of stability is problematic for identifying requirement-relevant conversation turns (Step \#1) and filtering based on Q\&A structures (Step \#2), where consistency is critical. Traditional ML models, in contrast, provide reliable, repeatable classifications once trained. Second, we conducted a preliminary experimentation where we established the feasibility of using LLMs for the first two steps of our approach: the results showed that LLMs often hallucinate, fabricating details or misidentifying conversational structures---a report of these analyses is available in our online appendix \cite{appendix}. This poses a significant risk in requirements engineering, where extracting information exactly as stated by stakeholders is essential. Errors in these early steps may compromise the integrity of the entire pipeline. In addition, incorporating LLMs in classification and filtering would have significantly increased the complexity of our study. Unlike structured ML models, LLMs require careful prompt engineering and tuning, making systematic evaluation difficult. Finally, recent research \cite{cheng2025generativeairequirementsengineering} confirmed that LLMs have not been extensively applied to conversational requirements engineering, meaning their effectiveness in these tasks remains largely unexplored. Given our focus on developing and assessing \textsc{RECOVER} as a structured approach, we opted for more controlled and grounded methods in the first two steps, leveraging LLMs only in Step \#3, where their generative capabilities were more appropriate.} The following sections outline the key design and implementation decisions behind our approach.

\subsection{Step \#1: Classification of Requirements-Relevant Conversation Turns}
\label{sec:step1}
Stakeholder conversations often reveal valuable information, but they can also introduce ambiguity \cite{ferrari2016ambiguity}. When manually eliciting requirements from these conversations, requirements engineers may become confused by ambiguous or noisy data, increasing the effort needed to complete the task. The first step of \textsc{RECOVER} aims at mitigating these challenges by employing advanced natural language processing techniques to automatically classify individual conversation turns as requirements-relevant or not. 

\smallskip
\textbf{Step \#1 Design.} This step reduces noise in stakeholder conversations by serving as a conversational \emph{filter} that refines dialogue turns. It begins by extracting and isolating individual snippets from each conversation turn. Using word-embedding techniques, the approach generates an $N$-dimensional representation of each turn \cite{levy2015improving}. This representation enables the extraction of features from the text, which natural language models can then use for classification. By transforming the conversation into real-valued vectors, this step prepares the data for subsequent analysis stages. 

These transformed sentences are analyzed by a machine learning (ML) model, which can binary-classify conversation turns as \emph{``requirements-relevant"} or \emph{``non-requirements-relevant"}. Our approach is intended to support requirements engineers in the elicitation phase: we indeed envision the output of the approach to be manually inspected by an expert. As such, we prioritize minimizing missed requirements over incorrectly including irrelevant turns, i.e., based on the expected use case of the approach, missing requirements in conversation turns would be much worse than having turns that do not contain requirements. Consequently, a good ML algorithm employed in this step should aim at optimizing \emph{recall} over \emph{precision}.

The output of this first step is represented by a list of individual requirements-relevant conversation turns. 

\smallskip
\textbf{Step \#1 Implementation.} From an operational standpoint, implementing the first step required identifying a suitable dataset to train the requirements-relevant conversation turns classifier. Since no existing dataset contains information linking conversation turns to related system requirements, we sourced data from existing natural language tasks in the requirements engineering field. Among the available datasets, the PURE dataset \cite{ferrariPURE} is one of the most widely used. It contains 79 publicly available natural language requirements documents collected from the Web, totaling 34,268 sentences. We used an annotated version of this dataset from Ivanov et al. \cite{ivanov2022extracting}, which includes 7,745 sentences from requirements specification documents labeled as \emph{``requirement"}  or \emph{``not requirement"}.  By training on this data, the model employed in Step \#1 can predict whether an individual conversation turn may contain a requirement.

To determine the optimal word embedding technique and machine learning model to use, we conducted preliminary experimentation. For word embeddings, we tested five different techniques: \textsc{TF-IDF} \cite{vanrijsbergen1979information}, \textsc{BERT} \cite{devlin2019bert}, \textsc{Word2Vec} \cite{mikolov2013efficient}, \textsc{FastText} \cite{bojanowski2016enriching}, and \textsc{GloVe} \cite{pennington-etal-2014-glove}. For the machine learning model, we experimented with 27 algorithms provided by the \textsl{Lazy Predict} library.\footnote{\textsl{Lazy Predict}: \url{https://github.com/shankarpandala/lazypredict}} This Python library facilitates model comparison without requiring manual parameter tuning. Each machine learning algorithm was evaluated in conjunction with each of the word embedding techniques to identify the most effective combination for classifying requirements-relevant conversation turns. 

\begin{table}[!h]
\centering
    \caption{Step \#1: Configuration of requirements-relevant conversation turns classifier coming from our preliminary experimentation.}
    \label{table:final_configuration}
    
    \begin{tabular}{|l|c|} 
    \rowcolor{black!80}
    \multicolumn{2}{l}{\color{white}\textbf{SVC Hyperparameters}} \\\hline
    C   &   1\\\hline
    \rowcolor{gray!20}
    Gamma   &   100\\\hline
    Kernel    &  RBF\\\hline
    \rowcolor{black!80}
    \multicolumn{2}{l}{\color{white}\textbf{FastText + SVC Evaluation Metrics}} \\\hline
    \rowcolor{gray!20}
    {Precision}   &  {0.814539}\\\hline
    {Recall}   &  {0.865608}\\\hline
    \rowcolor{gray!20}
    {Accuracy}   &  {0.846090}\\\hline
    {F1-Score}   &  {0.839133}\\\hline
    \end{tabular}
\end{table}

Each sentence was initially represented using one of the embedding techniques. We employed ten-fold cross-validation \cite{kohavi1995study} to partition the dataset and evaluate each algorithm's performance using \textsl{Lazy Predict}. To efficiently handle the computational demands of testing various combinations of word embeddings and machine learning techniques, sentences were represented as fixed vectors of 100 tokens, irrespective of requirement size or embedding method. This approach minimized computational time while maintaining the integrity of the results. We assessed the performance of 135 combinations of word embedding and machine learning models using well-established metrics such as \emph{F1-score} and \emph{accuracy} \cite{ml_metrics}. The analysis identified the best model for each word embedding technique, which was then subjected to hyperparameter tuning using the \textsc{GridSearchCV} algorithm from \textsc{Scikit-learn}. This step ensured the reliability of the \textsl{LazyPredict} results by performing another round of cross-validation to validate and consolidate the initial findings.

The complete results of this preliminary analysis are available in our online appendix \cite{appendix}. According to our findings, the best balance between performance and efficiency was achieved using \textsc{FastText}\footnote{Available at: \url{https://fasttext.cc/}}, a library for efficient learning of word representations, as the word embedding technique, and \textsc{SVC} \cite{burges1998tutorial}, Scikit-learn's implementation of the Support Vector Machine algorithm, as the classifier. The hyperparameter configuration and evaluation metrics for this combination are detailed in Table \ref{table:final_configuration}. This solution proved satisfactory due to its high \emph{recall} performance. \revised{However, while this suggests that a portion of potentially requirement-relevant conversation turns is not identified, this does not necessarily imply that these are lost. It is important to clarify that the classification step does not operate directly on fully formed requirements but rather on conversation turns that may contain requirement-relevant information. Since not all conversation turns potentially relevant will necessarily lead to a concrete system requirement, the missing turns represent a worst-case scenario rather than an exact measure of missed requirements.}



\subsection{Step \#2: Processing of Conversation Turns}
The requirements-relevant conversation turns identified in the first step may lack sufficient context, i.e., earlier references that are necessary for clarity. The second step of \textsc{RECOVER} addresses this issue by cleaning the conversation turns: it removes those that do not provide enough detail to elicit requirements and refines those that do by enhancing them with additional context and clarity.


\smallskip
\textbf{Step \#2 Design.} The approach first analyzes the length of the requirements-relevant conversation turns. If these turns are shorter than a certain threshold, they are discarded. The rationale behind this filtering is that excessively short turns are unlikely to contain detailed or meaningful information necessary for accurately capturing system requirements. Short turns often lack specificity, making them less useful in providing the comprehensive insights needed for requirements elicitation. By focusing on longer conversation turns, the approach ensures that only those segments that are more likely to contain substantial and relevant information are considered, thereby attempting to enhance the overall quality and relevance of the extracted requirements.

Subsequently, the approach leverages the findings of Spijkman et al. \cite{reconsum_spijkman_2023}, who discovered that requirements information is often embedded within question-and-answer exchanges in conversations. These interactions typically offer sufficient context to clarify the scope and specifics of the requirements being discussed. Building on this insight, our approach identifies conversation turns containing questions and merges them with the subsequent turns, assuming that answers follow questions directly in stakeholder discussions. \revised{This design choice was motivated by two key considerations. First, ensuring traceability in conversational requirements engineering remains an open challenge, making it difficult to determine whether consecutive questions or responses refer to the same requirement. Instead of grouping entire discussion segments, we opted for processing question-answer pairs individually to avoid missing information. Second, segmenting loosely structured discussions would require advanced discourse analysis techniques, such as dialogue act classification \cite{reithinger1997dialogue} or contextual dependency tracking \cite{zhang2021context}, which remain underexplored in this domain. Given these challenges, our method balances completeness and simplicity, ensuring relevant requirements are extracted without excessive complexity.} This refinement enriches the requirements-relevant conversational turns by organizing them into a question-and-answer format. As an output, the approach provides a refined list requirements-relevant conversation turns.

\smallskip
\textbf{Step \#2 Implementation.} When implementing this approach, we filtered out conversation turns that were shorter than seven words. We chose this threshold based on experimental observations. Through testing different thresholds in the empirical evaluation (see Section \ref{sec:method}), we found that a minimum length of seven words is generally needed to capture complete thoughts or questions. This ensures that the conversation turns we keep are more likely to contain useful and actionable insights for requirements elicitation. 

To refine the conversation turns, the approach identifies question-and-answer patterns using a dialogue acts methodology. Specifically, we adopted the design by Spijkman et al. in \textsc{REConSum} \cite{reconsum_spijkman_2023} and utilized the DialogTag implementation,\footnote{Available at: \url{https://github.com/bhavitvyamalik/DialogTag}.} which employs a BERT \cite{devlin2019bert} architecture for classification. Our implementation labels speaker turns as questions when DialogTag identifies them as either \emph{``question"} 
or \emph{``or-clause"} sentence types. We then merge each identified question with the subsequent turn, ensuring that the context and continuity of the conversation are preserved.

\subsection{Step \#3: Generation of System Requirements}
In the final step, the approach processes a refined list of conversation turns. Although the previous steps have removed irrelevant conversation turns, those reaching this step may still include some noise due to the unstructured nature of stakeholder interactions, such as ambiguous phrasing or incomplete thoughts. As such, the generation process must be addressed by advanced techniques to effectively manage and interpret the diverse content.

\smallskip
\textbf{Step \#3 Design.} To tackle the final challenges and generate system requirements, \textsc{RECOVER} leverages a large language model, aligning with current trends in software engineering that utilize LLMs for various automated tasks, such as code generation and explanation \cite{fan2023large}. With these models' advanced text-generation capabilities, our approach generates system requirements from conversation turns, aiming to produce output that is accurate and contextually relevant. Upon completion, the approach aggregates the requirements generated for each requirements-relevant conversion turn, hence providing the requirements engineer with a final list of system requirements.

\smallskip
\textbf{Step \#3 Implementation.} \textsc{RECOVER} utilizes \emph{Llama 2} \cite{llama2}, a collection of pre-trained and fine-tuned generative text models available in scales from seven billion to 70 billion parameters, as the core engine for the generation process. The decision to use this LLM is based on two key considerations: it is freely available and can be executed locally. As such, it allowed us to keep control over computational resources while maintaining the flexibility to tailor the model to our specific needs. We selected the seven-billion-parameter version of Llama 2. 
Our approach applies the \emph{few-shot} prompting technique \cite{llama2, fewshot}, which involves providing examples in the prompt to guide the model toward better performance. We designed a two-shot prompt, i.e., with two examples, one positive and one negative. The former is an example of a requirements-relevant conversation turn, while the latter shows an example of a conversation turn containing no requirements. These serve as a template for the desired format of the responses, using boilerplate text rather than realistic examples.

\revised{This decision was made after extensive experimentation with different prompt formulations. For brevity, we provide a detailed report of these experiments, along with the rationale behind discarding each tested prompt, in our online appendix \cite{appendix}. To address the inherent variability of LLM outputs, we executed each prompt at least 10 times on the entire conversation, treating each run as an independent evaluation to assess the consistency of the generated results. While these 10 runs were executed to select the most suitable prompt, the final implementation of our approach only executes the prompt once. The primary criterion for selecting the final prompt was its ability to produce structured and actionable system requirements that adhered to the intended response format, as assessed in our evaluation (see Section \ref{sec:method}). The evaluation of LLM outputs for each prompt was conducted manually, ensuring that the generated requirements aligned with stakeholder discussions without introducing speculative or inferred content.}  

\revised{Our final prompt explicitly requests system requirements, emphasizing functional aspects while avoiding speculative generation of non-functional requirements. In early experiments, available in our online appendix \cite{appendix}, broader prompts that encouraged the extraction of user, component, or quality requirements led to inconsistent outputs. When asked to generate user or component-level requirements, the LLM often mixed high-level goals with detailed implementation assumptions, reducing the clarity of the extracted requirements. Similarly, prompts requesting non-functional requirements frequently resulted in hallucinated quality attributes that were not discussed in the conversation. Given our focus on extracting only explicitly stated requirements, we constrained the prompt to system requirements, ensuring alignment with engineering best practices and minimizing inference errors.}  
For the sake of clarity, the prompt used for the generation of system requirements is the following:  

\examplebox{Prompt used in \textsc{RECOVER}}{
Given the following excerpt of a conversation, derive the system requirements, if any.\\Please answer only with the list of derived requirements as output, as shown in the following examples.\\Please do not consider the example excerpts provided in the examples in your final answer.\\---- Example 1: ---- \\Example excerpt: Excerpt of conversation containing system requirements\\Example output: \\1. The system must have example feature X;\\2. The system must have example feature Y;\\---- Example 2: ---- \\Example excerpt: Excerpt of conversation that does not contain system requirements \\Example output: \\None}

Finally, the approach compiles the generated outputs into a comprehensive list of system requirements.


\steattentionbox{\textbf{Tool availability}. All the code, data, and performed experiments are available in our online appendix \cite{appendix}.}

    

\begin{figure}
    \centering
    \includegraphics[width=0.75\linewidth]{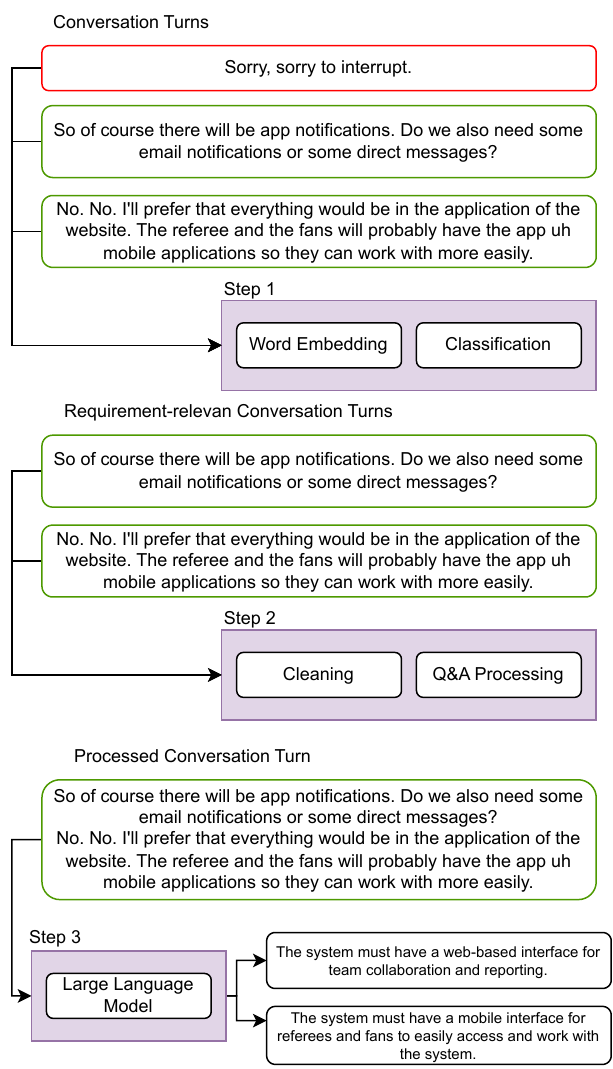}
    \caption{Running example of RECOVER.}
    \label{figure:example}
\end{figure}

For the sake of clarity, Figure \ref{figure:example} illustrates a practical instance of our framework, illustrating the process of extracting requirements from a conversation turn. This example showcases a successful classification derived from the empirical study discussed later in the paper. The conversation analyzed is an example conversation provided by Spijkman et al. \cite{reconsum_spijkman_2023}. Specifically, it highlights the outcomes when configuring the framework with the components discussed in the previous section, i.e., (1) \textsc{FastText} as the word embedding technique, (2) \textsc{SVC} for the binary classification task, (3) \textsc{Llama2} as the Language Model for the generation of requirements, and (4) a few-shot prompt for guiding the Language Model. As shown in Figure \ref{figure:example}, \revised{the approach first correctly identified two conversation turns as containing requirements-relevant information while discarding the irrelevant one. Then, \textsc{RECOVER} recognized the first turn as a question and linked it to the subsequent turn, which served as its answer. Finally, with both the question and answer providing context, the LLM in Step 3 generated a coherent and accurate set of system requirements}.

\begin{figure*}
    \centering
    \includegraphics[width=.8\linewidth]{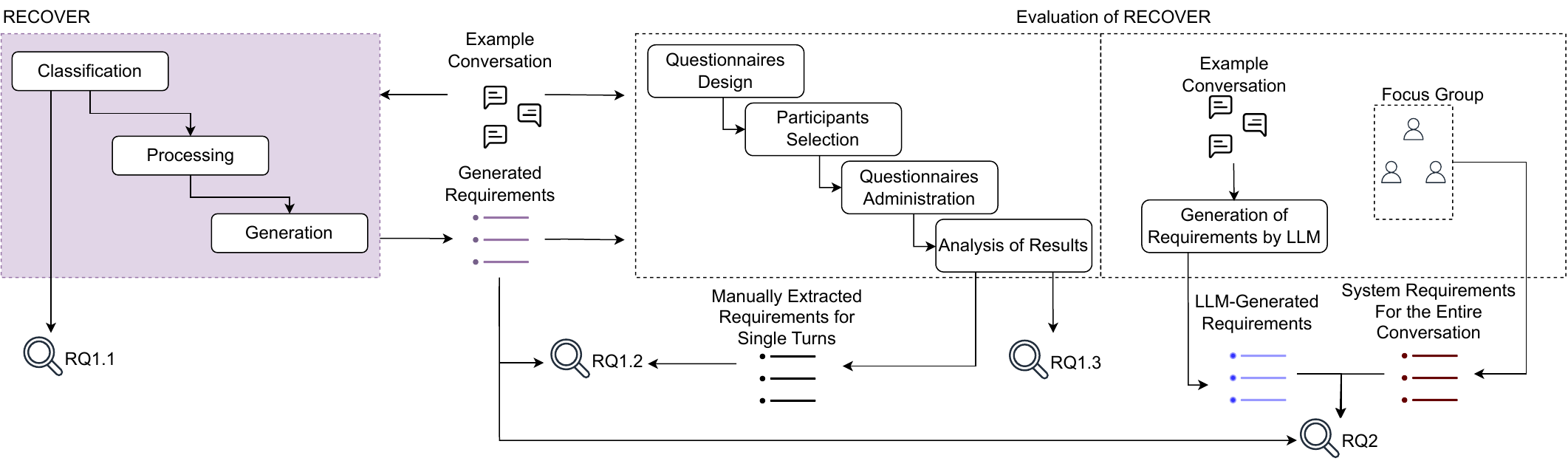}
    \caption{Overview of the research method proposed for our study.}
    \label{figure:method}
\end{figure*}

\section{Research Questions and Methods}
\label{sec:method}
The \emph{goal} of the study was to assess the extent to which system requirements may be automatically extracted from unstructured conversations between stakeholders using \textsc{RECOVER}, with the \emph{purpose} of supporting requirements engineers during the requirements elicitation phase. The \emph{perspective} is of both researchers and practitioners. The former are interested in assessing the feasibility of designing novel instruments to assist practitioners during requirements elicitation. The latter are interested in verifying the capabilities of an automated conversational requirements engineering instrument to assess its potential usefulness in a real-world scenario. To reach our goal, we designed a mixed-method study, as reported in the following.

\subsection{Research Questions and Context}
We organized the evaluation of \textsc{RECOVER} around two main research questions, covering two levels of granularity. The former assesses our approach at the level of \textbf{individual conversation turns} and aims at measuring the accuracy of the steps performed by the approach. This involves evaluating how effectively \textsc{RECOVER} identifies requirements-relevant conversation turns and generates accurate system requirements from them. Hence, we asked:

\sterqbox{RQ\textsubscript{1} – Individual Conversation Turns Assessment}{To what extent can the steps of \textsc{RECOVER} identify and generate high-quality system requirements from stakeholders' conversations turns?}

Addressing \textbf{RQ$_1$} involved conducting two main types of empirical evaluations. First, a statistical analysis to evaluate the accuracy of classifying requirements-relevant conversation turns from actual stakeholder documentation. Second, an analysis of the generated requirements, which was conducted through a user study, as this aspect required qualitative evaluation by experts to assess the relevance and quality of the outputs. For this reason, we split \textbf{RQ$_1$} according to the specific target of the evaluation. We empirically evaluated the classification to accomplish Step \#1 of \textsc{RECOVER} as part of the following sub-research question:

\sterqbox{RQ\textsubscript{1.1} – Requirements Classification}{To what extent can \textsc{RECOVER} \textit{identify} system requirements in stakeholders' conversations turns? }

Secondly, we evaluated the \textsc{RECOVER}'s ability to generate system requirements, hence covering Steps \#2 and \#3 of \textsc{RECOVER}, through the following sub-research question:

\sterqbox{RQ\textsubscript{1.2} – Requirements Generation}{To what extent can \textsc{RECOVER} \textit{extract} system requirements from stakeholders' conversations turns similarly to requirements engineering experts?}

Besides evaluating how similarly \textsc{RECOVER} can perform compared to requirements engineers, we also assessed the overall quality of the system requirements generated by considering three key attributes: \emph{correctness}, \emph{completeness}, and \emph{actionability}. \revised{Correctness refers to the accuracy of the requirements in reflecting the stakeholders' intentions, ensuring that they faithfully represent the information provided in the conversation without distortion or misinterpretation. This is a well-established criterion for assessing requirement quality \cite{3c, srsquality_mund_2015} and is distinct from completeness: while correctness evaluates whether the extracted requirements accurately capture what was stated, completeness assesses whether all relevant information has been included. Completeness evaluates whether all necessary requirements are captured, ensuring that no critical aspects are overlooked. This is a fundamental aspect of requirements engineering, as missing requirements can lead to incomplete or inadequate system implementations that fail to meet stakeholder needs \cite{3c, alvarez2002tell}. Actionability assesses how well the extracted requirements are structured to guide subsequent development efforts effectively. Automated approaches may produce requirements that lack specificity or clarity, potentially increasing the effort required for refinement before integration into the engineering process. Ensuring actionability helps minimize ambiguity and makes the generated requirements more practical for real-world use.} This research question covers Steps \#2 and \#3 of \textsc{RECOVER}, thus providing an additional overview of the actual capability of the approach:

\sterqbox{RQ\textsubscript{1.3} – Perceived Quality of Requirements}{What is the \textit{quality} of the system requirements generated in terms of correctness, completeness, and actionability?}


The second research question assesses our approach at the level of \textbf{entire conversation}. This assessment is crucial for determining the overall completeness of the requirements elicited by \textsc{RECOVER}. The approach aggregates requirements generated through a turn-by-turn analysis, which poses the risk of missing interdependencies or broader requirements that emerge only when considering the conversation as a whole. By focusing on the entire conversation, we aim to measure the extent of this risk:  

\sterqbox{RQ\textsubscript{2} – Entire Conversation Assessment}{What is the quality of the whole set of requirements generated by \textsc{RECOVER} from a stakeholders' conversation?}

The \emph{context} of the study is represented by one of the stakeholder's conversation\footnote{Available at: \url{https://github.com/RELabUU/REConSum/tree/main/data}} used by Spijkman et al. \cite{reconsum_spijkman_2023} to assess the performance of \textsc{REConSum}. This conversation is a stakeholders' dialogue transcript provided by Dalpiaz et al. \cite{dalpiaz_2021_conceptualmodel}, consisting of 133 speaker turns. \revised{The discussion revolves around the elicitation of requirements for a fictional software system designed for the International Football Association (IFA), where stakeholders explore functionalities to support teams, referees, and league managers in organizing and managing football leagues.} \revised{The transcript originates from a controlled experiment explicitly conducted for research in conversational requirements engineering, ensuring that the discussion reflects realistic stakeholder interactions rather than an artificially constructed dialogue.} \revised{Additionally, this conversation has been used in prior research as a reference point \cite{reconsum_spijkman_2023}, reinforcing its suitability as a benchmark for assessing automated requirements extraction approaches.} \revised{With its structured nature and well-defined scenario, the transcript provides a meaningful test case for demonstrating the capability of \textsc{RECOVER} to process stakeholder discussions and extract requirements effectively.} \revised{Table \ref{table:conversation} summarizes the characteristics of the conversation.}

\begin{table}[h]
\centering
    \caption{\revised{Characteristics of the conversation used to evaluate \textsc{RECOVER}}.}
    \label{table:conversation}
    
    \begin{tabular}{|l|c|} 
    \rowcolor{black!80}
    \multicolumn{2}{l}{\color{white}\textbf{Example Conversation }} \\\hline
    Domain &   Football \\\hline
    \rowcolor{gray!20}
    System's Name &   International Football Association (IFA)\\\hline
    Total Turns    &  133\\\hline
    \rowcolor{gray!20}
    Duration  &  39 minutes and 40 seconds\\\hline
    Provenance & Controlled Experiment\cite{dalpiaz_2021_conceptualmodel} \\\hline
    \end{tabular}
\end{table}

As part of the empirical assessment, we used this conversation as input for \textsc{RECOVER} to evaluate its performance and address the research questions. Figure \ref{figure:method} overviews the research methods employed to address the objectives of the study, while the remainder of the section details the specific research methods applied to each research question. In terms of reporting, we employed the guidelines by the \textsl{ACM/SIGSOFT Empirical Standards}.\footnote{Available at: \url{https://github.com/acmsigsoft/EmpiricalStandards}.} Given the nature of our study we followed the \textsl{``General Standard''}, and the \textsl{``Questionnaire Surveys''} guidelines.


\subsection{Research Methods for \textbf{RQ$_{1.1}$}}
To evaluate the capabilities of \textsc{RECOVER}'s Step \#1 in identifying requirements-relevant conversation turns, we first created an oracle consisting of a set of conversation turns labeled as either requirements-relevant or not. The oracle creation was performed by the first two authors of the article, who \emph{independently} analyzed the input conversation \cite{reconsum_spijkman_2023}. Specifically, they individually reviewed the conversation turns, using their expertise to determine whether each turn contained requirements-relevant information. To mark a conversation turn as requirements-relevant, they primarily exploited key indicators such as the presence of stakeholder needs, system functionalities, constraints, or explicit references to design or implementation details. After completing their independent assessments, they engaged in a discussion to compare the labels assigned and resolve any discrepancies that emerged from their individual evaluations.

The output of this process served as an oracle against which \textsc{RECOVER}'s predictions were compared. Turns identified as containing requirements and correctly predicted by the approach were marked as \emph{True Positives (TP)}. Conversely, turns not containing requirements but incorrectly predicted as containing them were marked as \emph{False Positives (FP)}. Similarly, turns predicted as not containing requirements were labeled as \emph{True Negatives (TN)} if they were indeed requirement-free, or \emph{False Negatives (FN)} if they actually contained requirements. We finally addressed \textbf{RQ$_{1.1}$} by computing \emph{precision} and \emph{recall} of the predictions.

\subsection{Research Methods for \textbf{RQ$_{1.2}$}-\textbf{RQ$_{1.3}$}}
\label{sec:resmetRQ1.2.3}
To evaluate the efficacy of the LLM component in \textsc{RECOVER}, we conducted surveys with experts in requirements engineering. We chose this survey-based approach due to the absence of a predefined oracle that specifies the actual requirements for each conversation turn identified in the first step. Through the survey, we addressed both \textbf{RQ\textsubscript{1.2}} and \textbf{RQ\textsubscript{1.3}}. On the one hand, the insights gathered allowed us to generate oracles for each conversation turn, allowing us to evaluate our approach's effectiveness compared to human-generated requirements. On the other hand, the survey had the goal to collect expert opinions on the quality of the requirements generated by our approach, enabling a more comprehensive assessment of its effectiveness.


\smallskip
\textbf{Questionnaire Design.} The questionnaire was structured as follows. First, participants were provided with an informed consent form that detailed the purpose of the study, the voluntary nature of their participation, and the confidentiality of their responses. They were informed that their expertise would be used to evaluate the performance of \textsc{RECOVER}, and that their input would be anonymized and used solely for research purposes. Participants were also made aware that they could withdraw from the study at any time without any repercussions. By proceeding, participants acknowledged their understanding of these conditions and consented to contribute their expertise to our research. 

Next, the survey presented a series of conversation turns and asked participants to elicit system requirements from them in an open-ended format. After completing this task, the survey displayed \revised{for each conversation turn a set of system requirements that could be elicited from them}---these were generated by \textsc{RECOVER}, although this was not disclosed to the participants. Participants were then asked to evaluate the quality \revised{each individual set of requirements extracted by RECOVER for the corresponding conversation turns} using a Likert scale ranging from 1 (poor quality) to 5 (excellent quality), specifically assessing three quality attributes such as:

\begin{itemize}
    \item \emph{Correctness} - \revised{The requirements accurately reflect the stakeholders' intentions, ensuring that they faithfully represent the information discussed in the conversation without distortion or misinterpretation. Additionally, correct requirements adhere to established standards in Requirements Engineering, such as the IEEE System Requirements Specification standard \cite{ieee_830}}

    \smallskip
    \item \emph{Completeness} - The requirements capture all relevant information discussed in the conversation turn, ensuring that no critical aspects are overlooked. 

    \smallskip
    \item \emph{Actionability} - The requirements are clearly defined and practical, meaning they can be readily used for follow-up design or implementation purposes.

\end{itemize}

At the end of the questionnaire, participants were informed that the proposed requirements had been generated by an automated approach and were invited to share their perceptions of the outcomes. They were asked to reflect on the effectiveness of the automated process in generating accurate, complete, and actionable requirements compared to traditional, human-led elicitation methods. Additionally, participants were encouraged to provide feedback on any limitations or potential improvements they observed in the automated approach, as well as their overall confidence in using such a tool in real-world scenarios.

Before distributing the survey, we conducted a pilot test with a researcher and software engineering practitioner from our network. This pilot allowed us to refine the clarity of several questions, such as reiterating the conversation turns and providing clearer explanations of the terms correctness, completeness, and actionability. 

\smallskip
\textbf{Questionnaire Dissemination.}
The input conversation consisted of 57 conversation turns. Including all these turns in a single questionnaire would have been not only excessively time-consuming but also likely to cause participant fatigue, potentially compromising the quality of their responses. To mitigate these risks, we split the conversation into five segments, each containing 10 to 12 conversation turns, resulting in the creation of five distinct questionnaires. Afterwards, we recruited 20 professional requirements engineers from our contact network, i.e., we applied a convenience sampling strategy \cite{robinson2014sampling} to reach practitioners with experience in requirements elicitation and analysis ranging between 5 and 10 years. Convenience sampling was selected over other recruitment strategies, e.g., survey dissemination over social networks \cite{robinson2014sampling}, because of our willingness to maintain control over the study's participants, aiming for a smaller yet more focused qualitative study. This decision allowed us to work closely with experienced practitioners in requirements elicitation and analysis, ensuring a depth of insight and control that might have been challenging with a larger, less targeted survey approach. The participants were instructed on the tasks to accomplish through email and were given 15 days to return the survey. 

This procedure resulted in four participants for each survey, meaning that each set of conversation turns was evaluated by four different practitioners. This was a critical aspect of our study, as different requirements engineers might interpret the same turn in varied ways. By having multiple experts assess the same turns, we were able to capture a wider range of interpretations, ensuring that the variability in expert judgment was thoroughly considered in our analysis. All in all, the four different manually elicited sets of requirements for each conversation turn were used to address \textbf{RQ\textsubscript{1.2}}. Additionally, we collected 228 evaluations---four for each of the 57 turns---on the quality attributes discussed in \textbf{RQ\textsubscript{1.3}}. Finally, to further enrich our discussion and gain insights into the implications of our work, we collected 20 expert opinions on the potential of \textsc{RECOVER}.
 
All the questionnaires and the answers, alongside a report describing the participants' demographic, are available in our online appendix \cite{appendix}.

\begin{figure}
    \centering
    \includegraphics[width=0.7\linewidth]{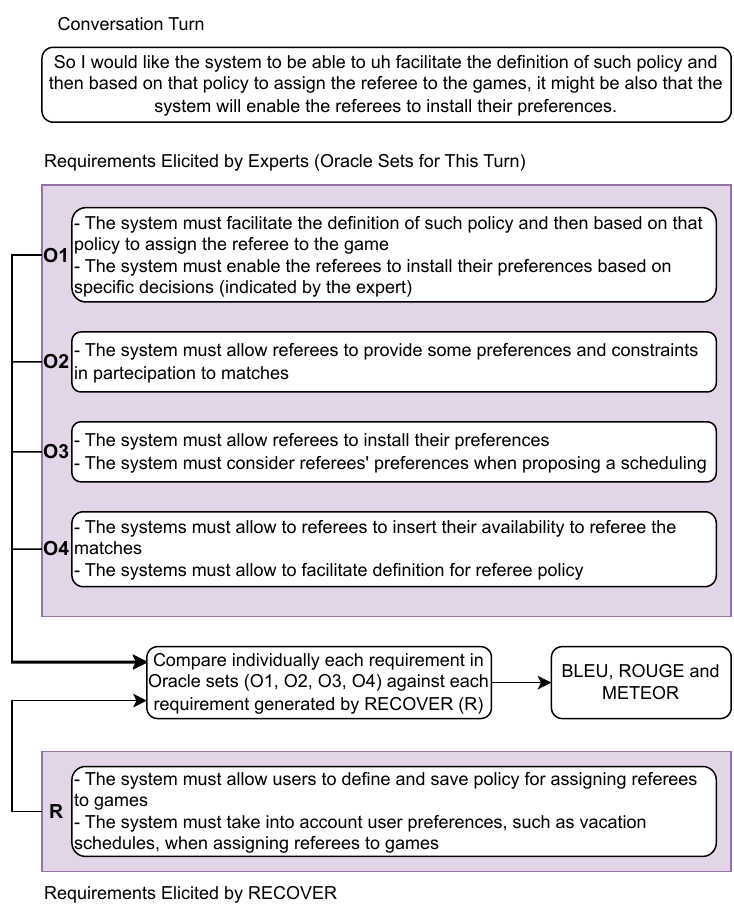}
    \caption{Example of the pairwise comparisons performed to answer RQ\textsubscript{1}.}
    \label{figure:pairwise}
\end{figure}

\smallskip
\textbf{Data Analysis.} As for \textbf{RQ$_{1.2}$}, we performed a quantitative evaluation using BLEU (Bilingual Evaluation Understudy) \cite{papineni2002bleu}, ROUGE (Recall-Oriented Understudy for Gisting Evaluation) \cite{lin-2004-rouge}, and METEOR (Metric for Evaluation of Translation with Explicit Ordering) \cite{banerjee-lavie-2005-meteor} scores to compare the generated requirements with the oracles---those generated by experts. These evaluation metrics are well-known and commonly used to assess the similarity between generated and reference text, providing values indicating the closeness of the machine-generated output to professional human-produced requirements. The BLEU score primarily emphasizes precision, evaluating the extent to which the generated text contains phrases present in the reference text. In contrast, ROUGE prioritizes recall by assessing how well the generated content covers the reference text. METEOR, instead, provides a more balanced evaluation by considering both precision and recall, thus offering a comprehensive analysis of the alignment between the generated text and the reference material. \revised{For each conversation turn included in our questionnaire, we obtained four distinct sets of requirements, each reflecting a different expert’s perspective. These sets served as independent oracles for evaluating RECOVER’s performance. Unlike a merged reference set, we retained each expert-provided set separately.} \revised{Given that \textsc{RECOVER} and the experts might articulate requirements in different but equivalent ways, we did not assume a strict one-to-one correspondence between individual requirements. Instead, each requirement extracted by \textsc{RECOVER} was compared against all requirements in the corresponding expert-provided oracle. This approach ensures that variations in phrasing do not artificially lower the similarity scores while still capturing how well \textsc{RECOVER}’s output reflects the intended requirements. Rather than selecting a single ``corresponding'' requirement from the oracle, we performed pairwise comparisons against all expert-provided requirements, computing similarity scores for each pair. \newrevised{These multiple similarity values—one for each comparison—were then averaged to produce a final similarity score for the candidate requirements. This aggregated score was used in our evaluation metrics to reflect the overall alignment between \textsc{RECOVER}'s output and each expert’s oracle.} Figure \ref{figure:pairwise} shows an example of this process: for each of the four oracles provided by experts, we compute the three metrics to assess the similarity between each requirement of the oracle against each requirement generated by \textsc{RECOVER} for the same conversation turn.}

As for \textbf{RQ$_{1.3}$}, we analyzed the responses from the questionnaire, focusing on participants' opinions regarding the three quality aspects considered, i.e., \emph{correctness}, \emph{completeness}, and \emph{actionability}. These aspects were rated on a Likert scale in the questionnaire. The responses were aggregated and summarized using frequency plots, which allowed us to visually represent the distribution of participants' ratings across the different quality attributes. This approach allowed us to identify how consistently the generated requirements were perceived by participants in terms of correctness, completeness, and actionability.

\subsection{Research Methods for \textbf{RQ$_2$}}
To evaluate the completeness and overall quality of \textsc{RECOVER} in extracting requirements from entire stakeholder conversations, we employed a similar approach to that used for the previous research questions. Specifically, we constructed an oracle based on the input conversation and compared \textsc{RECOVER}'s output against this oracle. Additionally, we compared the capabilities of our approach with a baseline LLM to assess the effectiveness of the various steps within \textsc{RECOVER}. This comparison aimed to determine the extent to which the structured, multi-step process of \textsc{RECOVER} provides tangible benefits over a more straightforward LLM-driven approach.

\smallskip
\textbf{Oracle Construction.} In the context of \textbf{RQ\textsubscript{1.2}}, we engaged requirements engineers to elicit an oracle of requirements from the input conversation. However, the requirements were extracted based solely on the specific conversation turns provided to the engineers, without considering the broader context of the entire conversation. This approach, while useful for evaluating \textsc{RECOVER} on the granularity of individual turns, may not capture the full scope and interconnectedness of the requirements that emerge when considering the entire conversation. Therefore, it was essential to assess how \textsc{RECOVER} performs when analyzing the conversation as a whole, ensuring that the approach can effectively integrate and synthesize information from multiple turns to generate comprehensive requirements. 

For this reason, we opted for the construction of a new oracle. To this aim, we involved three requirements engineering experts with professional experience ranging from 7 to 25 years. These experts currently work in industry and bring extensive practical knowledge of requirements elicitation and analysis. They were recruited from our contact network via email. We individually provided each expert with the input conversation and an Excel sheet structured into two columns: (1) the first, titled \textsl{`Conversation reference'}, was for indicating the specific parts of the conversation that informed the requirement; (2) the second column, titled \textsl{`Requirement'}, was for articulating the corresponding requirement in natural language. We did not impose a specific template for the experts to use when expressing the requirements, opting instead to give them the freedom to articulate the requirements in their preferred format. The experts were given 15 days to complete the task and return the filled Excel sheet via email.

After reviewing the three Excel sheets, we noticed that each expert approached the task differently. One expert generated a highly granular list, mapping nearly every sentence in the conversation to a requirement, resulting in 104 requirements. In contrast, the other two experts produced 36 and 17 requirements, respectively, as they often combined related points in the conversation or judged more sentences as not containing a requirement. Based on these observations, we organized a panel with the experts to consolidate their individual lists into a single, agreed-upon set of requirements. To guide the discussion and facilitate consensus, we employed the well-known Delphi method \cite{delphi}, which involves iterative rounds of anonymous feedback and discussion. This method allowed the experts to refine their assessments and converge on a final set of requirements through a structured process of evaluation and revision. In our case, the process lasted two rounds. In the first, the experts discussed their initial lists and, through iterative feedback, reduced the combined set to 40 requirements. In the second round, further refinement and discussion led to a final set of 28 consolidated requirements. 

These 28 system requirements formed the oracle through which we addressed \textbf{RQ$_2$}. When comparing this oracle with the one built through the turn-by-turn elicitation process, we found that the oracles were mostly similar. This similarity suggests that the turn-by-turn approach used by \textsc{RECOVER} represents a viable solution for automating the requirements elicitation process. However, we also noted significant differences, particularly in the depth of contextual integration and the capacity to capture comprehensive requirements spanning multiple conversation turns. For instance, one of the requirements on which the experts agreed was:

\begin{quote}
    \emph{``The system must allow any user (also not registered to the platform) to see and download information about events posted, ensuring public access is read-only and including data export formats.''}
\end{quote}

In this regard, this information was not directly captured by \textsc{RECOVER}, as it spanned over multiple conversation turns. Instead, the automated approach generated three separate requirements, i.e., (1) the system must allow users to view information about events posted; (2) the system must enable users to download information about events posted; and (3) the system must ensure that public access to event information is read-only and includes support for data export formats. While related, these requirements missed the contextual cohesion that experts used to form a single requirement, other than additional possible pieces of information enclosed within the conversation, e.g., that all users, including those not registered to the platform, should have access to the features above.

These differences could impact the overall performance of the approach, justifying the need for further investigation in \textbf{RQ\textsubscript{2}}. The whole oracle construction process, along with the lists of requirements, are in our online appendix \cite{appendix}.

\smallskip
\textbf{Baseline Selection.} As a second step to address \textbf{RQ\textsubscript{2}}, we selected an LLM-driven baseline to assess the added value of \textsc{RECOVER}'s structured, multi-step approach compared to a direct application of a large language model. Among available options, we chose ChatGPT~\cite{chatgpt}, powered by GPT-4~\cite{Achiam2023GPT4TR}, due to its strong performance and widespread use in software and requirements engineering~\cite{hou2024largelanguagemodelssoftware}. GPT-4 was used only in \textbf{RQ\textsubscript{2}}, which evaluates \textsc{RECOVER}’s end-to-end ability to generate system requirements. This allowed us to test whether prompting an LLM with the entire conversation could match the results of our structured pipeline. In contrast, \textbf{RQ\textsubscript{1}} focuses on evaluating each step of the pipeline separately, and applying GPT-4 to each subtask would have required substantial additional experimentation, beyond the scope of this study.


From an operational standpoint, the alternative list of requirements was generated by prompting ChatGPT to extract system requirements from the provided conversation. The prompt instructed ChatGPT to carefully analyze the entire conversation, which was supplied as an external file, and identify key requirements based on the dialogue between stakeholders. We deliberately chose not to employ any prompt engineering techniques for two reasons: (1) to assess whether \textsc{RECOVER} could outperform a basic baseline, representing the lower bound of LLM performance---if \textsc{RECOVER} failed to outperform this baseline, then advanced prompt engineering would be unnecessary at this stage, as the inadequacy of our approach would already be established; (2) optimizing a prompt engineering technique would have required a separate experimental setup, which was beyond the scope of our study. 

\smallskip
\textbf{Data Analysis.} With the requirements extracted by \textsc{RECOVER} and the baseline in hand, we compared them against those in the expert-generated oracle. We used the same evaluation metrics as in \textbf{RQ\textsubscript{1.2}}, specifically BLEU \cite{papineni2002bleu}, ROUGE \cite{lin-2004-rouge}, and METEOR \cite{banerjee-lavie-2005-meteor}. Additionally, we included two further metrics: Brevity Penalty (BP) and Length Ratio (LR), both defined within the context of BLEU \cite{papineni2002bleu}. The inclusion of BP and LR is justified because we are comparing entire corpora of text, i.e., full lists of requirements, rather than individual requirements for each conversation turn. More specifically, in \textbf{RQ\textsubscript{1.2}}, \revised{the metrics were computed through pairwise comparisons of each requirement generated by RECOVER for a specific conversation turn with all the requirements in the set of requirements of the corresponding oracle}. However, in \textbf{RQ\textsubscript{2}}, we shifted to a corpus-level evaluation, assessing the entire set of requirements as a whole rather than evaluating each requirement individually.  The reason for this shift is twofold. First, the sizes of the lists to compare were different---135 requirements from \textsc{RECOVER}, 31 from ChatGPT, and 28 from their mutual agreement---and a requirement-by-requirement evaluation could lead to misleading results, as it would not account for the varying levels of granularity across the lists. Comparing mismatched numbers of requirements could skew the analysis, giving an inaccurate representation of each approach's effectiveness. Second, the goal of the evaluation was to capture how closely the entire set of requirements generated by our approach matched the oracle, making a corpus-level evaluation more appropriate. As such, BP and LR are particularly valuable as they provide insights into the conciseness and overall length of the generated requirements lists: BP assesses verbosity relative to the oracle, while LR indicates their proportional length.


\section{Analysis of the Results}
\label{sec:results}
The following sections discuss the results that addressed the two main research questions of the study. 

\subsection{\textbf{RQ\textsubscript{1}} - Individual Conversation Turns Assessment}
Our first research question assessed \textsc{RECOVER}'s capabilities in correctly identifying and generating high-quality system requirements from stakeholders' conversations. 

\subsubsection{\textbf{RQ\textsubscript{1.1}} - Requirements Classification}
Table \ref{table:rq1} overviews the results for \textbf{RQ\textsubscript{1.1}}. More specifically, the results provide an assessment of \textsc{RECOVER}'s performance in classifying conversation turns as either containing requirements (Req) or not containing requirements (NonReq). The analysis covered 133 conversation turns, with \textsc{RECOVER} predicting 62 of them as requirements-relevant and 71 as non-requirements-relevant. 

\begin{table}[h]
\centering
    \caption{Results achieved for \textbf{RQ\textsubscript{1.1}} (Req = Containing Requirements, NonReq = Non Containing Requirements).}
    \label{table:rq1}
    
    \begin{tabular}{|l|c|} 
    \rowcolor{black!80}
    \multicolumn{2}{l}{\color{white}\textbf{Count}} \\\hline
    Total Turns &   133\\\hline
    \rowcolor{gray!20}
    Predicted as Req    &  62\\\hline
    Predicted as NonReq    &  71\\\hline
    \rowcolor{black!80}
    \multicolumn{2}{l}{\color{white}\textbf{Evaluation}} \\\hline
    \rowcolor{gray!20}
    True Positive =  39  &  True Negative = 59\\\hline
    False Positive =  23  &  False Negative = 12\\\hline
    \rowcolor{gray!20}
    {Precision}   &  {0.6290322581}\\\hline
    {Recall}   &  {0.7647058824}\\\hline
    \end{tabular}
\end{table}

The evaluation of the model's predictions reveals that \textsc{RECOVER} correctly identified 39 turns as containing requirements (True Positives) and 59 turns as not containing requirements (True Negatives). This indicates that the model was fairly accurate in identifying both relevant and irrelevant turns. However, there were 23 instances where the model incorrectly identified a turn as requirements-relevant (False Positives) and 12 instances where it failed to recognize a turn that actually contained a requirement (False Negatives). These errors highlight areas where the model could benefit from further refinement, particularly in reducing the number of false positives. For instance, we may envision the experimentation of more advanced filtering mechanisms that leverage semantic analysis to better differentiate between turns that truly contain requirements and those that do not.

Despite a precision of around 63\%, the recall of the approach, at 77\%, demonstrates the model's effectiveness in identifying most of the relevant information, even though some relevant turns were still missed. As anticipated in Section \ref{sec:step1}, in our case we aim at favoring recall over precision: in a real-case scenario, the output of the approach is supposed to be manually reviewed by an expert; thus, it is far more important to capture as many relevant requirements as possible, even if it means including some irrelevant ones. Therefore, the higher recall rate suggests that \textsc{RECOVER} is effectively minimizing the risk of overlooking essential requirements, which aligns with the intended use case and makes the results satisfactory. 
Further refinement of the classification algorithm or the incorporation of additional contextual information would still be worth to improve the recall of the approach, e.g., implementing methods to dynamically adjust the sensitivity of the classifier based on the conversation's context might further reduce the likelihood of missing critical requirements.

\begin{findingbox1}
The classifier employed in Step \#1  of \textsc{RECOVER} provides valuable support in identifying requirements in real stakeholders' conversations, with a \emph{precision} of 63\% and, more importantly, a \emph{recall} of 77\%.
\end{findingbox1}

\subsubsection{\textbf{RQ\textsubscript{1.2}} - Requirements Generation}
\revised{Initially, the model classified 62 turns as containing requirements. Interestingly, \textsc{RECOVER} primarily extracted functional requirements, which accounted for 88\% of the total requirements generated by the approach. However, in 12\% of the total extracted requirements, \textsc{RECOVER} was also able to identify non-functional requirements when they were explicitly discussed by stakeholders. This is given by the design of the prompt employed in Step \#3: \textsc{RECOVER} was designed to extract system features explicitly discussed in conversations, avoiding the inference of non-functional requirements to prevent hallucinations. For example, our approach successfully identified NFRs like system scalability when discussed but did not infer them otherwise, confirming that its ability to capture NFRs depends on stakeholder input rather than model-driven assumptions. For the sake of brevity, we included an analysis of the NFRs elicited in our online appendix \cite{appendix}.}

\revised{A manual inspection revealed that 5 of these were misidentified, as they consisted of generic or filler utterances (e.g., 'All right, okay.' or 'Okay. Okay. Um.'). These turns were excluded from further evaluation, leaving 57 turns for expert review. Among these, 6 turns were determined by all experts to contain no requirements, contrary to RECOVER’s prediction. As a result, a total of 51 conversation turns were recognized as containing requirement-relevant information by both RECOVER and at least one expert.}

After accounting for these inaccuracies, the remaining 51 turns (84\% of the original 62) were validated by the experts, who successfully elicited requirements from them. This finding reinforces our results for \textbf{RQ\textsubscript{1.1}}, highlighting the model's general effectiveness while also identifying areas for improvement. \revised{We categorized the remaining 51 turns into four groups based on the number of experts who were able to recognize in them potential requirements. Particularly, participants' recognition of requirements was distributed as follows:}

\begin{itemize}
\item 31 out of 51 turns (60.78\%) were identified as requirements-relevant by all four participants (\revised{Group with 4 answers}).
\item 10 out of 51 turns (19.61\%) were identified as requirements-relevant by three participants (\revised{Group with 3 answers}).
\item 7 out of 51 turns (13.73\%) were identified as requirements-relevant by two participants (\revised{Group with 2 answers}).
\item 3 out of 51 turns (5.88\%) were identified as requirements-relevant by only one participant (\revised{Group with 1 answer}).
\end{itemize}

\begin{table}[ht]
\centering
\scriptsize
\setlength{\tabcolsep}{3.7pt}
\caption{Results achieved for \textbf{RQ\textsubscript{1.2}} in \%.}
\label{table:rq12}
\resizebox{\columnwidth}{!}{ 
\begin{tabular}{l|c|c|c|c|c|c|c|c|c}
\rowcolor{black!80}
&\multicolumn{3}{||c||}{\color{white}BLEU} & \multicolumn{3}{c||}{\color{white}ROUGE} & \multicolumn{3}{c}{\color{white}METEOR}\\
\hline
\hline
Group with & Min & Mean & Max & Min & Mean & Max & Min & Mean & Max \\
\hline
\hline
\rowcolor{gray!20}
1 answer & 14.8 & 14.8 & 14.8 & 32.66 & 32.66 & 32.66 & 46.71 & 46.71 & 46.71 \\
2 answers & 4.63 & 9.81 & 14.99 & 27 & 39.37 & 51.75 & 30.28 & 43.96 & 57.63 \\
\rowcolor{gray!20}
3 answers & 1.87 & 4.09 & 6.46 & 24.97 & 36.37 & 49.44 & 27.91 & 40.22 & 54.36 \\
4 answers & 1.91 & 3.91 & 7.18 & 28.37 & 39.6 & 53.92 & 29.41 & 41.46 & 57.98 \\
\hline
\hline
\rowcolor{gray!20}
Average & 3.03 & 5.39 & 8.56 & 27.77 & 38.53 & 51.49 & 30.25 & 41.87 & 56.56 \\
\end{tabular}
}
\end{table}

We then used the widely accepted metrics such as BLEU \cite{papineni2002bleu}, ROUGE \cite{lin-2004-rouge}, and METEOR \cite{banerjee-lavie-2005-meteor} to quantitatively evaluate the performance of \textsc{RECOVER} in alignment with expert opinions. Table \ref{table:rq12} reports the performance of \textsc{RECOVER} across different groups.

The BLEU scores are relatively consistent within each group, as evidenced by the minimal variance between the Min, Mean, and Max values. This suggests a stable performance in precision-oriented evaluation. \revised{However, the BLEU scores decline: as the number of participant-provided answers used for comparison increases, the scores tend to be lower.} 
This trend might indicate the system's troubles in maintaining precision with more complex or varied input data, highlighting a potential area for improvement in handling diverse or nuanced content. 

The ROUGE scores, focused on recall, are higher across all groups than BLEU scores, suggesting that the system is more effective in capturing the overall content of the reference material. An interesting observation is a progressive increase in ROUGE scores from ``1 answer'' to ``4 answers'', which might imply better performance in encompassing the key elements of the reference text with more diverse inputs.

METEOR scores, which balance both precision and recall, are consistently the highest among the three metrics. This indicates a strong overall alignment with the reference texts, considering both exact word matches and semantic similarity.
The METEOR scores are also consistent with the number of answers, contrary to the BLEU and ROUGE scores diverging between groups. This pattern suggests that \textsc{RECOVER} performs better in a comprehensive evaluation scenario, particularly regarding semantic understanding and paraphrasing. The average scores across all groups show that METEOR outperforms both BLEU and ROUGE, reinforcing that a more balanced evaluation metric can accurately represent the \textsc{RECOVER}'s capabilities.

\begin{figure*}
    \centering
    \includegraphics[width=0.65\linewidth]{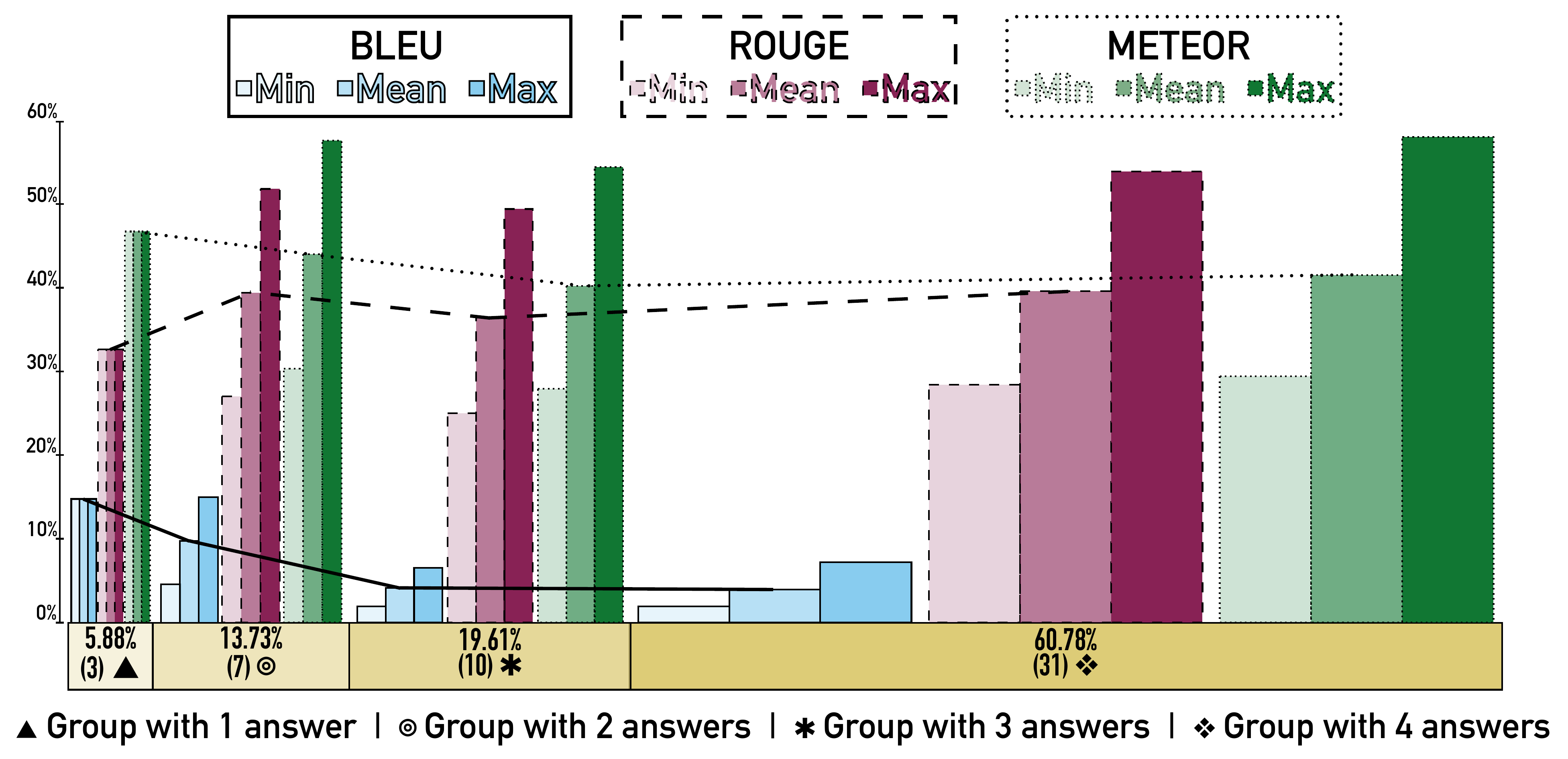}
    \caption{Results distribution in the different groups.}
    \label{figure:distribution}
\end{figure*}

As shown in Figure \ref{figure:distribution}, the pattern of increasing ROUGE and the consistent METEOR scores with the number of answers, contrasted with the decreasing trend of BLEU scores, underscores the importance of using multiple metrics for a well-rounded evaluation. It highlights the system's strengths in overall content capture and semantic understanding while pinpointing imprecision in exact wording. As such, the results underscore language processing tasks' complexity and multifaceted nature. All in all, while \textsc{RECOVER} shows promising capabilities in overall content understanding and semantic alignment, there is still room for improvement in matching the exact wording, especially in scenarios featuring more varied or complex inputs.

\begin{findingbox2}
The LLM employed in Step \#3 of \textsc{RECOVER} offers valuable assistance in generating requirements from real stakeholders' conversations, with a \emph{BLEU Mean score} of \emph{5.39\%}, a \emph{ROUGE Mean score} of \emph{38.53\%}, and a \emph{METEOR Mean score} of \emph{41.87\%}
\end{findingbox2}

\subsubsection{\textbf{RQ\textsubscript{1.3} }- Perceived Quality}
\begin{figure}
    \centering
    \includegraphics[width=.8\linewidth]{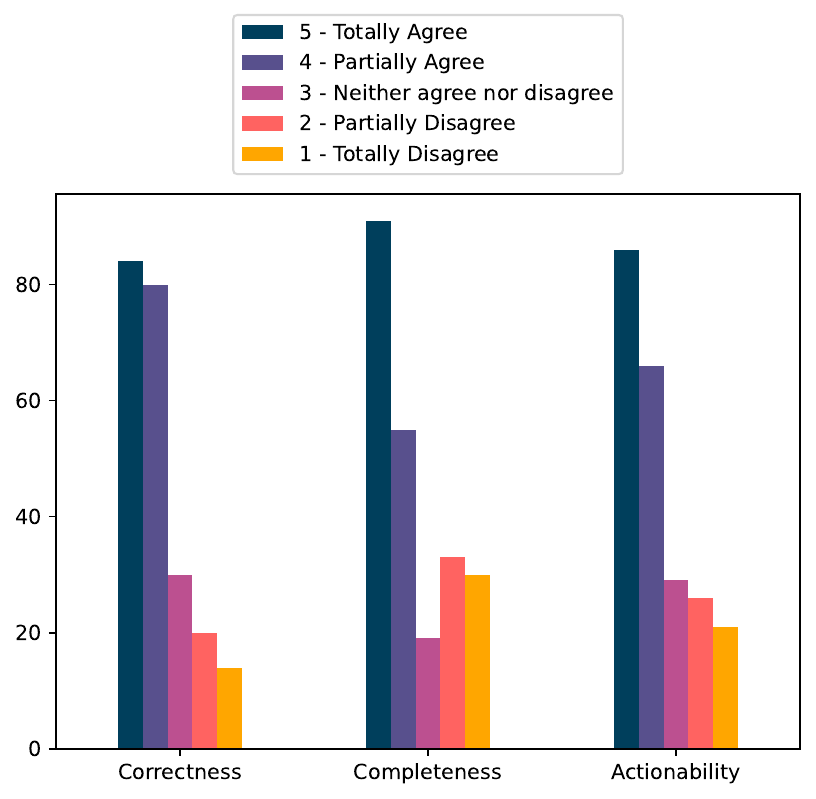}
    \caption{Questionnaire's answers: quality of the generated requirements.}
    \label{figure:rq3}
\end{figure}

To answer \textbf{RQ\textsubscript{1.3}}, we analyzed the experts' responses to our questionnaires. Participants rated the three quality indicators, i.e., (1) correctness, (2) completeness, and (3) actionability, on a Likert scale from 1 (poor quality) to 5 (excellent quality). Figure \ref{figure:rq3} shows the distribution of the answers. 

Considering total and partial agreement, 72\% of participants found \emph{correct} the generated requirements, 64\% believed that the requirements were \emph{completely} encapsulating what was discussed in the conversation turn, and 66\% considered the requirements as \emph{actionable}. Only 15\%, 21\%, and 20\% partially or totally disagreed with the generated requirements' correctness, completeness, and actionability, respectively. In the first place, these results indicate a generally positive reception of the requirements generated by \textsc{RECOVER}. The high agreement on the correctness of the requirements (72\%) reflects a broad consensus on their accuracy and alignment with the intended meaning of the conversation turns. \revised{Additionally, the majority of participants (64\%) agreed on the completeness of the requirements, suggesting that the approach is effective in capturing a substantial portion of the requirements discussed in stakeholder conversations.} 
Similarly, the strong agreement on actionability (66\%) indicates that most participants found the generated requirements to be practical and ready for implementation, further reinforcing the effectiveness.

In the second place, the relatively low percentages of disagreement indicate that, while a minority of participants identified some issues with the generated requirements, these concerns were not widespread. In this sense, the issues related to correctness, completeness, and actionability could likely be addressed by requirements engineers during their review of the output generated by our approach. The minimal occurrence of these issues suggests that \textsc{RECOVER} can still reduce the time and effort required to extract requirements from stakeholder conversations, allowing engineers to focus on fine-tuning and validating the results rather than starting from scratch. Based on these considerations, we may argue that our approach is able to produce requirements that are accurate, comprehensive, and actionable, though there remains room for improvement.

In the second place, the relatively low percentages of disagreement indicate that, while a minority of participants identified some issues with the generated requirements, these concerns were not widespread. In this sense, the issues related to correctness, completeness, and actionability could likely be addressed by requirements engineers during their review of the output generated by our approach. \revised{However, ensuring completeness requires more than just reviewing \textsc{RECOVER}’s output, as it also involves verifying that no critical information has been overlooked in the original stakeholder conversation, since some requirement-relevant conversation turns may not be identified. To address this, requirements engineers play a dual role: (1) refining and validating RECOVER’s output to ensure correctness, completeness, and actionability, and (2) cross-checking the original conversation to confirm that all essential information has been captured. This aligns with the iterative nature of requirements elicitation, where multiple refinements and discussions help mitigate the risk of overlooking key information.} However, the minimal occurrence of these issues suggests that \textsc{RECOVER} can still reduce the time and effort required to extract requirements from stakeholder conversations, allowing engineers to focus on fine-tuning and validating the results rather than starting from scratch. \revised{By explicitly reinforcing the necessity of human oversight, we align \textsc{RECOVER}’s intended role as a support tool rather than a fully autonomous requirement extraction system.} 
Based on these considerations, we may argue that our approach is able to produce requirements that are accurate, comprehensive, and actionable, though there remains room for improvement. In this respect, our participants provided insights into the key issues that may lead to further refinement of the approach. More specifically:

\noindent\textbf{Level of Detail.} Sometimes, the requirements are too specific, and it seems \textsc{RECOVER} hallucinates aspects of the conversation turn, e.g., it generates several times the same requirement or is not able to understand the actor of the requirement. In other cases, the requirements were too general, or perhaps no requirements should be elicited.

\noindent\textbf{Missing Context.} In general, it is hard to get the whole context by only analyzing a single conversation excerpt. \textsc{RECOVER} seems to be lacking the ability to detail actors and context-related aspects of requirements, yet we think the nature of the task emphasizes this issue.

The analysis of the final section of the questionnaire revealed some key insights worth highlighting. These questions explored participants' perceptions of using an automated approach for eliciting requirements from conversations. Since participants were initially unaware that the requirements presented were generated by an automated tool, the first question asked whether they still stood by their previous assessments of the quality of the generated requirements. Interestingly, 100\% of participants confirmed their original evaluations, reinforcing the initial positive assessment of the requirements generated by our approach. 

The second question, which provided more interesting insights, asked whether participants would consider adopting the automated approach that produced these requirements in their own work. In response, 65\% of participants said \emph{Yes}, 10\% said \emph{No}, and 25\% were undecided. Those who were unsure expressed concerns about the maturity of the approach, noting that they would currently use it only as a supportive tool rather than a primary method. These concerns are reasonable and perfectly align with the intended goal of our work. The purpose of \textsc{RECOVER} is indeed not to replace the role of requirements engineers but to serve as an aid in the elicitation process, enhancing efficiency and reducing manual effort. 

Finally, in the last question, we asked participants if they would use \textsc{RECOVER} to save time and effort. While some participants argued that the quality of the output is not yet sufficient to justify the trade-off with time and effort savings, 80\% still answered \emph{Yes}, indicating a strong willingness to integrate the tool into their workflows, provided that the output meets certain quality thresholds.

\begin{findingbox3}
Most participants (72\%) found the generated requirements \emph{correct}, 64\% reported that they \emph{completely} encapsulated what the conversation turn analyzed discussed, and 66\% considered the requirements as \emph{actionable}. Hence, we claim that requirements generated by \textsc{RECOVER} are of good quality, considering experts' opinions. We also highlight additional potential improvements that may further enhance the usefulness of the approach, ensuring that it continues to evolve in alignment with the needs of requirements engineers.
\end{findingbox3}

\subsection{\textbf{RQ\textsubscript{2} }- Entire Conversation Assessment}
Table \ref{table:rq2} presents the results of our evaluation of \textsc{RECOVER} at the level of entire conversations, comparing its performance with that of ChatGPT. The findings show that \textsc{RECOVER} achieved the highest BLEU (78.82\%) and METEOR (39.09\%) scores when compared to the oracle, indicating a high degree of precision (BLEU) and a strong alignment with the reference requirements in terms of content, wording, and structure, despite some variations in exact phrasing (METEOR). In contrast, ChatGPT recorded the highest ROUGE score (43.96\%), suggesting that it captured the most content similar to the oracle, as evidenced by a higher n-gram overlap. However, ChatGPT's lower BLEU score of 49.36\% and significantly lower BP (0.76) and LR (0.78) scores indicate that its generated requirements were shorter and less comprehensive compared to the oracle. These lower BP and LR scores suggest that, while ChatGPT effectively captures a broad range of content (as reflected in its high ROUGE score), it falls short in matching the exact length and depth of the oracle’s requirements, potentially sacrificing detail and completeness. 

These results suggest that \textsc{RECOVER} is more effective at generating detailed and accurate requirements that align closely with the reference standard, making it a more reliable approach for capturing the full scope and context of stakeholder conversations. As such, we could confirm that the multi-step approach employed by our approach supports the generation of higher-quality requirements compared to the simple use of a large language model.

To further understand the differences between \textsc{RECOVER} and the baseline, we compared the outputs of both approaches using the evaluation metrics. The results showed moderate performance across the board, with perfect Brevity Penalty (BP = 1) and Length Ratio (LR = 1) scores, indicating that the generated requirements were well-balanced in terms of conciseness and length. This suggests that the comprehensive steps within \textsc{RECOVER} applied to individual conversation turns do not result in any significant loss of information compared to using ChatGPT on the entire conversation. 

\begin{table}[ht]
\centering
\scriptsize
\caption{Results achieved for \textbf{RQ\textsubscript{2}} in \%.}
\label{table:rq2}
\resizebox{\columnwidth}{!}{ 
\begin{tabular}{l||c||c||c||c||c}
\rowcolor{black!80}
&{\color{white}BLEU} & {\color{white}ROUGE} & {\color{white}METEOR} & {\color{white}BP} & {\color{white}LR}\\
\hline
\hline
\rowcolor{gray!20}
RECOVER vs Oracle & \textbf{78.82} \faAngleDoubleUp & 37.20 & \textbf{39.09} \faAngleUp & 0.92 & 0.93 \\
ChatGPT vs Oracle & 49.36 & \textbf{43.96} \faAngleUp & 35.34 & 0.76 & 0.78\\
\hline
\hline
\rowcolor{gray!20}
RECOVER vs ChatGPT & 57.83 & 37.83 & 32.76 & 1 & 1 \\
RECOVER vs Expert 1 & 37.24 & 48.18 & 39.92 & 1 & 1\\
\rowcolor{gray!20}
RECOVER vs Expert 2 & 92.34 & 28.61 & 40.22 & 1 & 1 \\
RECOVER vs Expert 3 & 30.11 & 26.13 & 28.57 & 0.92 & 0.93 \\
\hline
\end{tabular}
}
\end{table}

To better understand how the different styles and granularities used by requirements engineers in eliciting requirements from the input conversation impacted the performance of \textsc{RECOVER}, we conducted an additional analysis, which involved comparing the output of our approach against the individual lists of requirements provided by each engineer before the application of the Delphi method.

When comparing \textsc{RECOVER} with Expert 1, the approach demonstrated high ROUGE (48.18\%) and METEOR (39.92\%) scores, indicating strong content similarity and alignment with Expert \#1's elicited requirements. This suggests that \textsc{RECOVER} was particularly effective in capturing the breadth of content identified by Expert \#1 and maintaining a close semantic relationship with the expert's phrasing. The comparison with Expert \#2 yielded exceptionally high BLEU (92.34\%) and METEOR (40.22\%) scores, suggesting that \textsc{RECOVER} closely mirrored the precision and recall of Expert \#2's requirements. These scores indicate that \textsc{RECOVER} was able to reproduce Expert \#2's requirements with minimal deviation in wording and content, reflecting a high degree of accuracy and detail in the requirements generated by the tool. However, the results were different when compared to Expert \#3. \textsc{RECOVER} displayed lower scores across all metrics, suggesting less similarity and alignment with Expert \#3's elicited requirements. This discrepancy may be attributed to differences in the granularity and style used by Expert \#3, which \textsc{RECOVER} might not have captured as effectively as it did with the other experts. These observations suggest that the overall performance of our approach is not necessarily influenced by the number of requirements to be elicited. If the number were the determining factor, we would have expected to see significant differences in the comparison with Expert \#1, who provided the largest number of requirements (104). However, the major differences in performance were observed when comparing \textsc{RECOVER} with Expert \#3, who provided the fewest requirements (17). This suggests that the level of granularity, with Expert \#3 adopting a more abstract and concise approach, may have influenced the overall similarity, as \textsc{RECOVER} may have struggled to fully capture the broader, less detailed requirements. \revised{These findings reinforce the inherent subjectivity in requirements elicitation, where differences in interpretation, assumptions, and granularity naturally lead to variation among experts.} \revised{To further investigate this phenomenon, we performed a manual analysis of the lists of requirements generated by \textsc{RECOVER} and by the three experts, where we found cases in which also the experts elicited different and contrasting requirements for the same aspect of the system. For instance, concerning the management of teams' budgets, Expert \#3 thought that the system should guarantee access to budget data to both administrative and managerial people from within the teams, while Expert \#2 only considered implementing a budgeting portal for "teams" in general. However, Expert \#1 elicited different requirements, in which he wrote that the system should allow the IFA administration---the owners of the system---to manage budget data. \textsc{RECOVER}, in contrast, elicited requirements that indicated that both the teams and the IFA administration should be allowed to manage the budget. Rather than reflecting a flaw in \textsc{RECOVER}, these differences highlight the challenge of defining a single ground truth in requirements extraction.}
As a consequence, further refinements could focus on enhancing the approach's ability to adapt to the specific style and level of detail employed by each requirements engineer, ensuring that \textsc{RECOVER} effectively supports varying methodologies and granularity in requirements elicitation.

Additionally, a manual evaluation of the lists of requirements revealed that while the requirements captured by both experts and \textsc{RECOVER} generally encapsulated the same aspects of the conversation, there were notable differences in their structure and organization. The experts tended to organize their requirements around user roles and associated functionalities, addressing various aspects such as user interaction within the system. These requirements were detailed and segmented into specific sections based on the actors involved, providing a clear outline of user responsibilities and interactions. In contrast, the requirements generated by \textsc{RECOVER} were structured more around specific features and processes rather than user roles. The list was segmented by functional areas, with a strong emphasis on financial aspects and less focus on user experience. The structure was more modular, centering on discrete functionalities within the system. For example, consider the following requirement elicited by Expert \#1:
\emph{``The system must allow the IFA representative to schedule and reschedule games, with an interface for scheduling and ensuring notifications for changes.''}
As shown, the requirement explicitly refers to the user who must be supported. On the contrary, \textsc{RECOVER} expressed the same requirement in terms of the features the system should provide. In particular:
\emph{``The system must have a scheduling portal to manage team schedules and events.''}

Overall, the requirements produced by the experts were more detailed, which was expected. However, we observed that most of the experts' requirements could be mapped onto subsets of those generated by \textsc{RECOVER}. In some instances, the requirements generated by our automated approach were repetitive, reflecting slight variations stemming from different conversation turns, while the experts were able to consolidate similar information into more cohesive, well-formed single requirements. In any case, our results show that \textsc{RECOVER} is capable of capturing a broad range of relevant information from stakeholder conversations, effectively generating requirements that align closely with experts. Despite some redundancy, the approach provides a solid foundation for requirements elicitation, possibly reducing the manual effort required by experts while still allowing them to refine and consolidate the output as needed. 


\begin{findingbox4}
The BLEU (78.82\%) and METEOR (39.09\%) scores indicate that \textsc{RECOVER} is highly effective at generating precise and comprehensive requirements. While ChatGPT's higher ROUGE score (43.96\%) suggests it captures a broad range of content, its lower BLEU (49.36\%) and BP/LR values (0.76 and 0.78) point to issues with precision and completeness. This comparison underscores \textsc{RECOVER} as a more reliable tool for producing requirements that closely align with oracle's standards.
\end{findingbox4}

\begin{table*}[!ht]
\centering    
    \footnotesize
    \caption{\newrevised{Results of our \emph{``in-vivo''} evaluation on three industrial meetings conversation transcripts. Step 1 - Negative values refer to the conversation turns classified as not requirement-relevant by \textsc{RECOVER}. Steps 1 and 2 - Positive refers to the conversation turn positively classified as requirement relevant after their processing in Step 2. Finally, Step 3 refers to metrics computed by analyzing the requirements generated for each turn. The metrics are the same used to answer \textbf{RQ\textsubscript{1.3}} and the value reported is the average of all conversation turns.}}
    \label{tab:invivoeval}
    \begin{tabular}{l||rr||rrrr||rrr} 
    \hline
    \rowcolor{black}
    & \multicolumn{2}{c||}{\textbf{\color{white}Step \#1 - Negative}} &\multicolumn{4}{c||}{\textbf{\color{white}Steps \#1 and \#2 - Positive}} & \multicolumn{3}{c}{\textbf{\color{white}Step \#3}} \\
    & \textbf{\% of TN} & \textbf{\% of FN} & \textbf{System} & \textbf{Technical} & \textbf{Management} & \textbf{Unrelated} & \textbf{Correct} & \textbf{Complete}  & \textbf{Actionable}  \\\hline 
     \rowcolor{gray!40}
    C1 - 271 valid turns & 83.85  & 16.15 & 44 (40\%) & 38 & 60 & 12 & 4.30 & 4.25 & 4  \\
    C2 - 143 valid turns& 91.94  & 8.06 & 28 (34\%) & 20 & 24 & 9 & 3.46 & 3.75 & 3.5  \\
    \rowcolor{gray!40}
    C3 - 165 valid turns & 93.81  & 6.19 & 5 (7\%) & 34 & 20 & 9 & 3.8 & 4 & 4.4 \\
    \hline
    \end{tabular}
\end{table*}
\section{\newrevised{Further Analyses and Implications}}
\label{sec:discussion}
\newrevised{The results discussed in Section~\ref{sec:results} demonstrate RECOVER's effectiveness in identifying and generating system requirements from stakeholders' conversations. Despite these promising findings, we acknowledge a potential threat to the transferability of the results. The evaluation was conducted using a \textbf{single conversation transcript} obtained from a controlled experiment specifically designed for conversational requirements engineering research~\cite{dalpiaz_2021_conceptualmodel}. At the time of our study, this was the only publicly available dataset to support an \emph{end-to-end evaluation} of our approach. This setup, while methodologically rigorous, reflects what may be considered an \emph{``in-vitro''} setting. In response to this potential limitation, we designed a complementary \emph{``in-vivo''} experiment based on independent stakeholder conversation transcripts drawn from a different, industrial elicitation scenario. This addition aims to assess the consistency of \textsc{RECOVER}'s behavior in a more naturalistic and noisy setting, mitigating concerns regarding the transferability of our findings to varied conversational contexts. As a consequence of the considerations above, this section is structured as follows. In the first part, we report the in-vivo evaluation of \textsc{RECOVER}, discussing the results and comparing them with those of the original evaluation. In the second part, we present the broader implications of our findings.} 

\newrevised{\subsection{RECOVER: An In-Vivo Evaluation}}
\newrevised{To conduct our in-vivo evaluation, we applied \textsc{RECOVER} to additional transcripts derived from real-world industrial meetings, as collected in prior work on requirements elicitation from natural language conversations~\cite{rodeghero2017userstories}. These conversations were recorded during regular stand-up meetings and teleconferences at a U.S.-based software company, where the first author of that study acted as an intern. The dataset comprises 27 anonymized conversations extracted from 9 meetings, spanning approximately 24 hours of recorded audio. While originally collected in the context of user story analysis, the conversations naturally reflect a broad spectrum of team interactions across the development lifecycle - which is expected, given their origin in routine stand-up meetings involving developers, project managers, and other stakeholders. This diversity makes the dataset particularly valuable, as it captures realistic and varied communication scenarios. At the same time, it calls for a careful selection of conversations most aligned with the focus of \textsc{RECOVER}, ensuring that the evaluation remains coherent with the tool's intended support for requirements elicitation.\footnote{\newrevised{The data were shared with us for research purposes, yet the original recordings and transcripts cannot be made publicly available due to privacy constraints and the sensitive nature of the discussions.}} Specifically, from the dataset we retained \emph{three} conversations based on their completeness, clarity, and topical coherence - these are referred to as Conversation C1, C2, and C3 for confidentiality reasons. These were among the most self-contained discussions, making them suitable for a meaningful end-to-end assessment with \textsc{RECOVER}. The remaining conversations were not retained as they primarily took place during later stages of the development and were predominantly focused on technical updates or managerial coordination (e.g., agile procedures, story point estimation, task scheduling). As such, they could not meaningfully support the evaluation of \textsc{RECOVER} in requirements elicitation, the primary scenario it is designed to assist.}

\newrevised{As an additional consideration, it is important to note that the nature of these conversations differs significantly from the one used in our main evaluation. Whereas our primary study focused on an \emph{initial stakeholder meeting}, hence representing an early-stage requirements elicitation, these additional transcripts reflect \emph{ongoing development contexts}. Despite this mismatch, we considered them valuable for three reasons. First, they offer insight into how \textsc{RECOVER} might perform when applied to ongoing development contexts, where requirements often evolve or emerge implicitly. Second, the conversations span different types of software systems, including web platforms and cyber-physical systems, thus providing a testbed for evaluating generalization across domains. Third, their relatively noisy and unstructured nature enables a more rigorous examination of Step \#1 of our pipeline, which aims to filter out non-functional turns and focus on requirement-relevant content.}


\newrevised{For each of the three conversations (C1, C2, and C3), we performed an initial cleaning to remove sentences that were non processable, i.e., those containing placeholder annotations used by the original authors to indicate unintelligible speech or content deemed irrelevant to the study context. After this cleaning, the total numbers of valid turns for each conversation were respectively 271, 143, and 165, as shown in Table \ref{tab:invivoeval}. Then, we executed \textsc{RECOVER} on each cleaned transcript and tracked its behavior across all three processing steps. Specifically, we recorded: (1) the turns discarded after Step \#1, having been classified as not containing requirement-relevant information; (2) the turns retained after Step \#1 and further processed by Step \#2, which maps utterances to specific requirements-related concepts; and (3) the final list of requirements automatically generated from each retained turn in Step \#3.}

\newrevised{Table~\ref{tab:invivoeval} reports the results of our analysis. To validate the effectiveness of \textsc{RECOVER}'s initial filtering (Step \#1), we manually inspected the discarded turns to estimate the proportion of false negatives, i.e., requirement-relevant turns that were incorrectly filtered out (column \textsl{``\% of FN''} in Table~\ref{tab:invivoeval}). This verification was conducted independently by two of the authors and resolved through discussion, following the same protocol adopted in our main evaluation, to mitigate subjectivity in judgment. Overall, \textsc{RECOVER} demonstrated a strong ability to correctly discard non-relevant content, with the percentage of true negatives (column \textsl{``\% of TN''}) ranging from approximately 84\% to 94\%. The observed false negative rates were relatively low (between 6\% and 16\%) and are largely influenced by the conversational context: as previously noted, these transcripts include discussions not strictly tied to system functionalities. This further reinforces the value of Step \#1 in managing noisy, real-world inputs. This is especially true when considering Conversation C3, where only 7\% of turns were related to system functionalities: also in this case, the false negative rate is below 7\%, indicating robust filtering even in challenging conditions.}


\newrevised{To further investigate the nature of the turns retained after Step \#1 and processed by Step \#2, we then conducted a manual annotation to categorize their content. As outlined earlier, this step targets turns that were considered potentially requirement-relevant and thus passed through the second stage of \textsc{RECOVER}, which maps utterances to specific requirements-related concepts. To mitigate subjectivity, the annotation was carried out independently by two of the authors and conform through discussion. Specifically, each turn was classified into one of four categories: \textsl{`System'} related, i.e., turns discussing system functionalities that could inform requirement elicitation; \textsl{`Technical'}, covering implementation or architectural details discussed by developers; \textsl{`Management'}, involving planning, scheduling, or coordination; and \textsl{`Unrelated'}, referring to conversational content not pertinent to the system or project. The distribution of these labels is reported in Table~\ref{tab:invivoeval} (column \textsl{``Steps \#1 and \#2 - Positive''}). The results confirm that only a small fraction of the retained turns directly relate to system functionalities: for instance, just 7\% in Conversation C3. This finding aligns with our earlier observation that most conversations in this dataset reflect later-stage project discussions. It is also consistent with prior analyses by Rodeghero et al. \cite{rodeghero2017userstories}, who reported that only 5.5\% of the turns in their dataset were relevant for the elicitation of functional requirements. These outcomes further underscore the importance of Step \#1 in filtering noise and reinforce the value of targeted approaches like \textsc{RECOVER} in extracting requirements-relevant content from broader project discussions.}


\newrevised{Finally, building on the turns identified as \textsl{System} related in the previous step, we evaluated the quality of the system requirements generated by \textsc{RECOVER} in Step \#3. For each requirement elicited from these turns, we manually assessed its quality using the same criteria adopted for \textbf{RQ\textsubscript{1.3}}: \textit{Correctness}, \textit{Completeness}, and \textit{Actionability}, each rated on a Likert scale from 1 (poor) to 5 (excellent). The evaluation was independently conducted by two of the authors and reconciled through discussion. The results, shown in Table~\ref{tab:invivoeval} (column \textsl{``Step \#3''}) provide further insights into \textsc{RECOVER}'s behavior in real-world conditions. Conversation C1, which had the highest number and proportion of turns related to system functionalities (40\% of the positives), yielded the strongest outcomes, with all quality indicators rated 4 or above. This confirms that when suitable content is available, \textsc{RECOVER} is capable of extracting high-quality requirements. In Conversation C2, the scores were slightly lower - ranging from 3.46 to 3.75 - which may be attributed to the smaller number of functional turns and a higher proportion of ambiguous or mixed-content utterances. Nevertheless, the extracted requirements still demonstrated acceptable quality. Interestingly, despite Conversation C3 having the fewest functional turns (only 7\%), the average quality ratings remained relatively high (between 3.8 and 4.4), suggesting that \textsc{RECOVER} can still produce valuable outputs even from limited and fragmented input.}

\newrevised{We acknowledge this in-vivo evaluation is preliminary; nevertheless, the analysis reinforces  \textsc{RECOVER}'s robustness across every stage—from filtering irrelevant content to generating coherent, actionable requirements—and indicates its promise for real-world development contexts.}


\newrevised{\subsection{Broader Discussion and Implications}}
The evaluation of \textsc{RECOVER} reveals several strengths and areas for improvement in automating the extraction of system requirements from stakeholder conversations. A key strength is its high recall rate (77\%), effectively capturing most relevant requirements and minimizing the risk of missing critical information, making it a valuable support tool for human experts who can further refine the output. Additionally, \textsc{RECOVER} demonstrates strong performance in generating precise and comprehensive requirements, as evidenced by its high BLEU and METEOR scores compared to expert-generated oracles.


However, it is important to clarify that recall in this context refers to the identification of requirement-relevant conversation turns rather than fully formed requirements. This means that while a portion of relevant turns were not identified, it does not necessarily equate to actual requirements being lost. Not all turns flagged as relevant by experts would have led to concrete system requirements, making this a worst-case estimate rather than an exact measure of missed content. Moreover, requirements elicitation is an iterative process, allowing engineers to revisit and refine extracted information, which further mitigates the impact of potential omissions. Nonetheless, assessing whether this level of recall is sufficient in practice would require further empirical investigations across different domains. \newrevised{Our in-vivo evaluation offers preliminary support for this, showing that \textsc{RECOVER} performs reliably across its pipeline even when applied to noisy, real-world conversations from industrial meetings. Despite differences in context, the tool consistently identified system-functional content and generated actionable requirements, suggesting potential for broader applicability beyond controlled settings.}

The approach also faces challenges. \textsc{RECOVER}’s emphasis on recall over precision introduces trade-offs, as capturing a broader set of potential requirements increases the risk of extracting irrelevant information. \revised{Striking the right balance between maximizing recall and minimizing noise remains a key challenge for practical adoption.} Additionally, while \textsc{RECOVER} outperforms the baseline, it sometimes struggles with maintaining context, leading to repetitive or overly granular requirements that lack the cohesion of those produced by experts. Future refinements should focus on improving contextual understanding and better adapting to different requirements engineering styles to enhance usability and effectiveness.

\steSummaryBox{\faHandORight \hspace{0.05cm} Take Away Message \#1.}{\textsc{RECOVER} shows promising results in eliciting actual requirements, effectively capturing the majority of relevant information. However, there is room for improvement in managing context and reducing the inclusion of irrelevant requirements.}

The emphasis on \textsc{RECOVER}'s understanding of the application domain and the importance of clearly defined business objectives signifies the pivotal role of domain expertise. Employing our approach in the preliminary phase for objective definition underscores a potential collaboration between human expertise and automated tools.

While participants noted advantages such as the generation of more specific and less ambiguous requirements, our study does not directly demonstrate that RECOVER reduces time and effort in requirements extraction. Although analysts qualitatively assessed the completeness and usefulness of \textsc{RECOVER}’s output, verifying automatically generated requirements may not necessarily require less effort than extracting them manually. Despite this, our results show that \textsc{RECOVER} may serve as a valuable supporting tool for practitioners, particularly in the early phases of requirement elicitation. \newrevised{The in-vivo evaluation reinforces this by showing that the modular structure of \textsc{RECOVER} helps manage noisy or fragmented input, increasing robustness in real-world settings and supporting practical human-in-the-loop use.} Therefore, while our findings suggest potential benefits, further research is needed to quantify the impact of RECOVER on efficiency and effort in real-world settings. 

\steSummaryBox{\faHandORight \hspace{0.05cm} Take Away Message \#2.}{\revised{The reported benefits of the approach emphasize the potential advantages of LLMs in improving both the efficiency and quality of the requirements engineering process, provided they are carefully customized.}}

The concern about generating irrelevant requirements underscores the need for a careful approach when integrating AI into the field of requirement engineering.

We believe our work lays the foundation for future research in conversational requirements engineering. Starting from conversations allows practitioners to uncover key system aspects early on, helping to address quality attributes like ethics, privacy, and fairness. Better understanding how requirements are discussed can lead to higher-quality solutions, particularly in security and ML-enabled systems.

\section{Threats to Validity}
\label{sec:ttv}
Although our study highlights that \textsc{RECOVER} may represent a valuable tool for requirements engineers, we are aware of the threats and limitations that must be discussed.

\smallskip
\textbf{Threats to Internal Validity.} To address \textbf{RQ\textsubscript{1.1}}, we evaluated the ability of Step \#1 of \textsc{RECOVER} to identify requirements-relevant conversation turns. Because of the lack of data in the field, we had to perform such an assessment by hand. Indeed, we recognize that there is a threat caused by potential biases or wrong labeling. However, to mitigate such an issue, the authors of this paper evaluated the conversation in different rounds and each separately, discussing the final results by combining each one's opinion.

Moreover, for \textbf{RQ\textsubscript{1.2}}, we evaluated the similarity of the requirements generated by \textsc{RECOVER} for each conversation turn against an oracle. The same approach was applied to answer \textbf{RQ\textsubscript{2}}, but in this case, the evaluation was based on an oracle derived from the entire conversation. These oracles were collected by surveying professional requirements engineers within our contact network. While we acknowledge potential subjectivity in the elicitation process, two key aspects give us confidence in the reliability of the oracles. First, the expertise of the professionals involved: all participants had substantial experience in requirements engineering, with backgrounds ranging from several years to decades in the field. Second, the rigorous research methods employed to synthesize the information gathered from these experts, such as the Delphi method used in \textbf{RQ\textsubscript{2}}, ensured a well-rounded and consensus-driven set of requirements.

\revised{Finally, \textsc{RECOVER} does not handle conflicting requirements, as it extracts requirements from individual conversation turns without analyzing the discussion holistically. Since conflict resolution requires human judgment \cite{van1998managing}, automating it without full context could lead to inaccurate decisions. Similarly, RECOVER lacks a traceability mechanism, which could help link extracted requirements to their original discussion but remains a challenge in conversational requirements engineering. Given these complexities, we designed RECOVER to assist rather than replace requirements engineers, though adding conflict detection and traceability is a promising future direction.}

\smallskip
\textbf{Threats to Construct Validity.} We developed a prototype implementation of the \textsc{RECOVER} framework by experimenting with diverse machine-learning algorithms for Step \#1, examining 135 combinations of ML algorithms and word-embedding techniques. Although advanced deep-learning models might have provided additional insights into our approach's capabilities, we prioritized examining simpler models due to their lower computational costs and increased interpretability. The good classification performance achieved by these simpler models has supported our confidence in our design choices. \revised{Acknowledging the potential value of exploring more complex models, we performed a preliminary analysis using GPT-4 to perform this task. While the LLM demonstrated a general ability to recognize feature-related discussions, we observed a critical issue: it often fabricated details, such as timestamps, or inconsistently split conversation turns. These hallucinations made it unsuitable for \textsc{RECOVER}’s structured approach, which required precise identification of requirement-relevant turns without distorting the original conversation. As part of our future work, we plan to further experiment with LLMs and prompt engineering to potentially improve our approach by incorporating more recent technology.}

For Step \#2, we implemented Question\&Answer tagging using DialogTag, based on the design of previous works \cite{reconsum_spijkman_2023}. Additionally, we made the simplifying assumption that each identified question was immediately followed by its corresponding answer in the next turn. While this assumption streamlines our approach, we acknowledge the potential limitations it introduces. \revised{To experiment with more powerful techniques, we performed a preliminary evaluation of GPT-4 also for this step of \textsc{RECOVER}. We prompted GPT to detect question-answer pairs and determine whether they contained requirement-related content. Similar to Step \#1, GPT struggled with consistency, frequently misidentifying Q\&A relationships or incorrectly discarding relevant information.} Our future work will aim at enhancing this step by incorporating conversation disentanglement techniques \cite{elsner2008you} to better handle more complex dialogue structures \revised{and experimenting with LLMs and prompt engineering}.

Additional threats arise from using a Large Language Model in Step \#3. We chose Llama2 as the primary engine for the generation process, but further experimentation with other LLMs is necessary to explore potential improvements. To minimize the risk of hallucinations, we carefully crafted our prompts to be formal and precise \cite{rawte2023exploring}. Furthermore, we tested several different prompts, which are available in our online appendix \cite{appendix}, and each was executed at least 10 times to ensure the reliability of the results. \revised{Despite our careful prompt design, the decision to explicitly request system requirements while excluding non-functional requirements introduces a potential threat. Specifically, this choice may limit the ability of \textsc{RECOVER} to capture quality-related concerns unless they are explicitly mentioned in the conversation. While this design minimizes hallucinations and ensures that extracted requirements are grounded in stakeholder discussions, it may overlook implicit requirements that stakeholders assume rather than state explicitly.}

\smallskip
\textbf{Threats to Conclusion Validity.} To mitigate potential threats of drawing conclusions from a questionnaire, we carefully designed it to minimize ambiguity and the potential for response biases. Moreover, we performed a pilot test to ensure the understandability and correctness of the questions. To avoid potential biases regarding using LLMs in SE tasks, we did not mention that the requirements shown were artificially generated until the very end of the questionnaire. Finally, we acknowledge that our work draws conclusions about the efficacy of the ML model employed in Step 1 by only considering trivial metrics such as precision and recall. Although we experimented with many different solutions, we know that further experimentation would be needed.

\smallskip \newrevised{\textbf{Threats to External Validity.} A potential threat to the generalizability of our findings lies in the use of a single stakeholder conversation for the primary evaluation of \textsc{RECOVER}. Although this transcript originates from a controlled experiment specifically designed for conversational requirements engineering research and has been adopted in prior work as a reference benchmark~\cite{dalpiaz_2021_conceptualmodel}, its scope remains limited to a single elicitation setting. Given that stakeholder conversations can vary widely across domains, team roles, and project phases, this may not fully reflect the diversity of real-world requirements engineering practices. To mitigate this limitation, we complemented the primary evaluation with an in-vivo analysis involving three additional conversations drawn from industrial development meetings. Although these conversations were not originally focused on early-stage elicitation, they exposed \textsc{RECOVER} to a broader range of communication scenarios, system types, and conversational structures. The results confirmed the pipeline’s robustness even in less structured, later-stage contexts, offering preliminary evidence of the approach’s adaptability to real-world conditions. Nonetheless, future work will extend this evaluation further by incorporating a wider and more representative set of stakeholder conversations to strengthen external validity.}

\section{Conclusion}
\label{sec:conclusion}
We proposed \textsc{RECOVER}, an approach designed to assist requirements engineers in generating system requirements from stakeholders' conversations. Our findings indicate that while \textsc{RECOVER} can reduce elicitation effort, it still benefits from human oversight for validation and refinement.

Future work will include broader experimentation with \textsc{RECOVER}, including larger-scale qualitative and industrial evaluations involving conversations from diverse contexts. We also aim to extend the approach by integrating further steps in the RE process, such as automatically generating user stories while ensuring mechanisms for human validation to maintain accuracy and reliability, and tailored prompting strategies to extract non-functional and user-centric requirements while minimizing hallucination risks. Additionally, we plan to explore traceability and conflict resolution techniques to better identify related or overlapping conversation turns. Finally, we will investigate fine-tuning and retrieval-augmented generation (RAG) to further improve the precision and contextual awareness of the generated requirements. These enhancements represent promising directions for advancing \textsc{RECOVER}.



\section*{Acknowledgments}
We acknowledge the support of the European Union - NextGenerationEU through the Italian Ministry of University and Research, Project PRIN 2022 PNRR ``FRINGE: context-aware FaiRness engineerING in complex software systEms" (grant n. P2022553SL, CUP: D53D23017340001). \newrevised{We thank the Associate Editor who handled our manuscript for their invaluable support, and the anonymous reviewers for their thoughtful feedback and constructive suggestions.}

\balance
\scriptsize
\bibliographystyle{IEEEtran}
\bibliography{references}

\IEEEpubid{\makebox[\columnwidth]{\hfill}}
\section*{Authors' Biographies}
\vspace{-1cm}

\begin{IEEEbiography}
[{\includegraphics[width=1in,height=1.25in,clip,keepaspectratio]{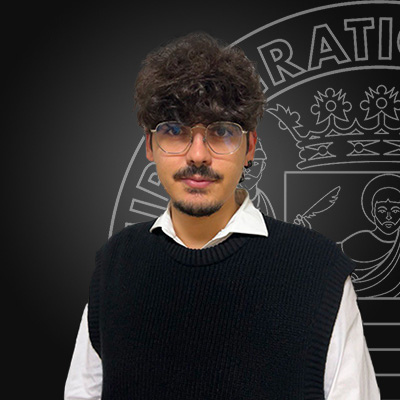}}]{Gianmario Voria} is a Ph.D. Student at the University of Salerno (Italy). His research covers software engineering for AI, with a particular focus on ethics and fairness, human and social aspects of software engineering, and empirical software engineering.
\end{IEEEbiography}

\vspace{-1.15cm}

\begin{IEEEbiography}
[{\includegraphics[width=1in,height=1.25in,clip,keepaspectratio]{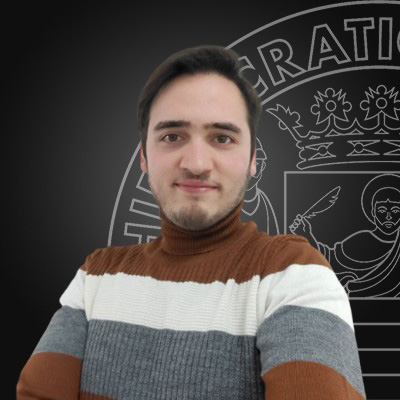}}]{Francesco Casillo} holds a Ph.D. in Computer Science from the University of Salerno, where he focused on Natural Language Processing techniques for the detection of non-functional requirements, including privacy, security, and fairness. His research spans requirements engineering and software maintenance, with a focus on the application of large language models.
He has presented his work at leading international venues in software engineering. 
\end{IEEEbiography}

\vspace{-0.75cm}

\begin{IEEEbiography}
[{\includegraphics[width=1in,height=1.25in,clip,keepaspectratio]{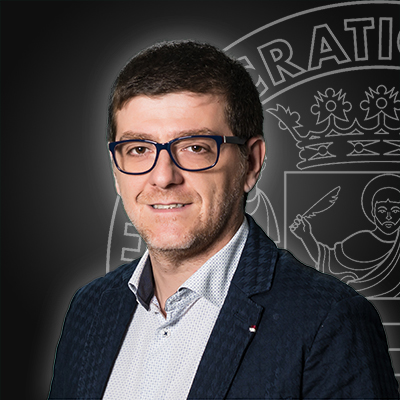}}]{Carmine Gravino} is a Full Professor at the University of Salerno, and co-director of the SQM/Web Engineering Lab. His research in Software Engineering spans visual modeling, machine learning-based prediction, search-based techniques, and software maintenance. More recently, he has focused on engineering educational metaverses and addressing non-functional requirements such as privacy and security. He has authored over 100 publications in international venues.
\end{IEEEbiography}

\vspace{-0.75cm}

\begin{IEEEbiography}
[{\includegraphics[width=1in,height=1.25in,clip,keepaspectratio]{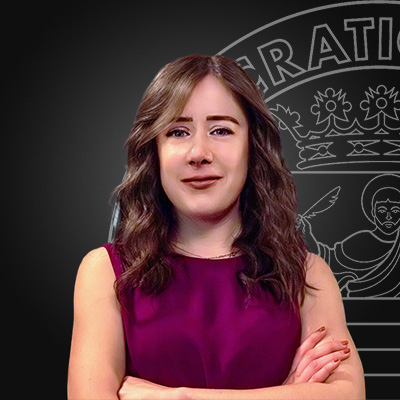}}]{Gemma Catolino} is an Assistant Professor at the Software Engineering (SeSa) Lab (within the Department of Computer Science) of the University of Salerno. Her research covers social software engineering, software maintenance and evolution, code quality, and empirical software engineering. She has co-chaired several workshops and conferences, served as co-general chair for MobileSoft 2024, and serves on leading SE committees and editorial boards. 
\end{IEEEbiography}

\vspace{-0.75cm}

\begin{IEEEbiography}
[{\includegraphics[width=1in,height=1.25in,clip,keepaspectratio]{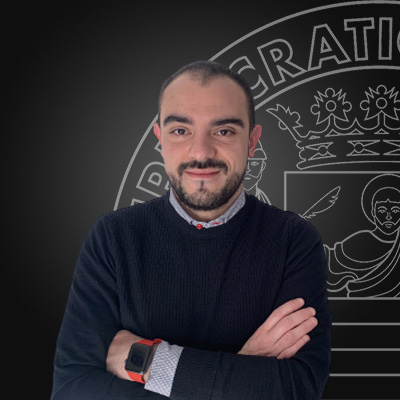}}]{Fabio Palomba}
is an Associate Professor of Computer Science at the University of Salerno (Italy). His research covers software maintenance, code quality, and empirical software engineering. He received the 2023 IEEE/TCSE Rising Star Award and multiple recognitions for papers and reviewing. He has co-chaired ICPC 2021 and SANER 2024, and serves on leading SE committees and editorial boards.
\end{IEEEbiography}

\end{document}